\documentclass[preprint,review,12pt]{elsarticle}

\let\today\relax
\makeatletter
\def\ps@pprintTitle{%
    \let\@oddhead\@empty
    \let\@evenhead\@empty
    \def\@oddfoot{\footnotesize\itshape
         { } \hfill\today}%
    \let\@evenfoot\@oddfoot
    }
\makeatother

\biboptions{sort&compress}

\usepackage{epsfig}
\graphicspath{{Figure/}}
\usepackage{booktabs}
\usepackage{natbib}
\usepackage{amssymb}
\usepackage{graphicx,epstopdf}
\usepackage[pdf]{pstricks}
\usepackage{subcaption}
\usepackage{amsmath}
\usepackage{float}
\usepackage{xcolor}
\usepackage{tabularx}
\usepackage{verbatim}

 \usepackage{lineno}

 \makeatletter
 \providecommand{\url}[1]{%
   \begingroup
     \let\bibinfo\@Secondoftwo
     \urlstyle{rm}%
     \href{http://dx.doi.org/#1}{%
       \discretionary{}{}{}%
       \nolinkurl{#1}%
     }%
   \endgroup
 }
 \makeatother

\usepackage{hyperref}
\hypersetup{
    colorlinks=true,
    linkcolor=blue,
    filecolor=blue, 
    citecolor=blue,
    urlcolor=blue,
    pdftitle={Manuscript},
    pdfpagemode=FullScreen,
    }

\begin{document}

\begin{frontmatter}



\title{Enhanced Profile-Preserving Phase-Field model of Two-Phase Flow with Surfactant Interfacial Transport and Marangoni Effects}


\author[inst1]{Haohao Hao}
\author[inst1]{Xiangwei Li}
\author[inst1]{Tian Liu}
\author[inst1]{Huanshu Tan\corref{cor1}}
\ead{tanhs@sustech.edu.cn}
\cortext[cor1]{Corresponding author}

\affiliation[inst1]{organization={Multicomponent Fluids group, Center for Complex Flows and Soft Matter Research \& Department of Mechanics and Aerospace Engineering, Southern University of Science and Technology},
            city={Shenzhen},
            postcode={518055}, 
            state={Guangdong},
            country={China}}

\begin{abstract}
Using a regularized delta function to distribute surfactant interfacial concentration can simplify the computation of the surface gradient operator $\nabla_s$, enabling the phase-field model to effectively simulate Marangoni flows involving surfactant transport.
However, the exact conservation of  total surfactant mass is compromised due to deviation from the equilibrium phase field profile, numerical diffusion, and mass non-conservation in each phase. 
To overcome these limitations, we have developed a new model for simulating two-phase flow with surfactant transport along the interface. 
This model employs a profile-preserving strategy to maintain the equilibrium interface profile, ensuring accurate calculation of the regularized delta function and better surfactant mass conservation. 
Within the framework of the advective Chan-Hilliard phase-field model,  we utilize a regularized delta function with a reduced gradient to minimize numerical diffusion. 
Furthermore, we introduce a hybrid surface tension model that integrates the free-energy and the continuum-surface force models to mitigate spatial discretization errors, particularly in scenarios with high density and viscosity ratio. 
Verification tests demonstrates the model's effectiveness in simulating surface diffusion on stationary and expanding drop, suppressing spurious currents, and capturing the deformation of two-dimensional drops in shear flow.
The results closely align with analytical solutions and previous numerical studies. 
Finally, we apply the model to investigate the contraction and oscillation dynamics of a surfactant-laden liquid filament, revealing the role of the Marangoni force in shaping filament behavior.

\end{abstract}


%


\end{frontmatter}

\section{Introduction}\label{sec:s1}

Surface-active agents, known as surfactant, possess a unique molecular structure that includes both hydrophilic (water-loving) and hydrophobic (water-repellent) groups, leading to their accumulation at fluid interfaces. 
As the interfacial surface tension depends on the surfactant concentration, a non-uniform distribution of surfactant creates variations in normal (capillary) and a tangential (Marangoni) stress along the interface. 
These surface tension variations significantly influence the dynamics of two-phase flows, including droplet migration~\citep{Banerjee2016,Tan2021,Banerjee2020}, drop deformation~\citep{Stone_leal1990,FEIGL20073242}, drop pinch-off~\citep{Roche2009}, wave breaking~\citep{Erinin_Liu_Liu_Mostert_Deike_Duncan_2023}, drop-interface coalescence~\citep{dong2023}, drop formation in a microfluidic channel~\citep{Kalli_2023, Anna2006}, and filament contraction~\citep{Kamat2020,Constante2020}.
This ability to modify interfacial properties makes surfactants widely applicable in various fields, including spraying and coating~\citep{Song2017}, self-propelled droplets~\citep{Michelin2023}, inkjet printing~\citep{detlef2022}, superhydrophobic drag reduction~\citep{Fernando2023}, and emulsion and oil recovery~\citep{Perazzo2018}.

To capture the distribution of surfactant concentration during the evolution of two-phase interface and reveal the hydrodynamics related to Marangoni effect, several numerical models have been developed by coupling the surfactant transport equation and the hydrodynamic equations. 
These models can be classified into two main categories, i.e., interface tracking models (Lagrangian description)~\citep{JAMES2004685,LAI20087279,MURADOGLU20082238,Muradoglu2014,Shin2018} and interface capturing models (Eulerian description)~\cite{XU20125897,CLERETDELANGAVANT2017271,STRICKER2017467,ERIKTEIGEN2011375,farsoiya2024coupled,Liu_2018,Liu_Zhang_Ba_Wang_Wu_2020}, which compute the evolving position of the interface and the surfactant concentration.
In interface tracking models, the surfactant transport equation~(Eq.~\eqref{STE_O1}) is discretized and solved on a Lagrangian grid, where the surfactants are directly tracked along the interface.
In contrast, interface capturing models can distribute surfactants along a diffuse interface region of finite thickness, implicitly capturing their behavior at the two-phase interface.
Compared to interface tracking approaches, interface capturing models are easier to implement, particularly when simulating interfacial flows with complex topological changes, such as breakup~\citep{Deike2022} and coalescence~\citep{Pirouz2015}.


One proposed distribution formulation that rearranges surfactant concentration over a diffuse interface is based on a regularized delta function, giving a surfactant transport equation (Eq.~\eqref{STE_O2}). 
This formulation avoids the difficulties associated with computing the surface gradient operator \(\nabla_{s}\), allowing for its application in various simulations~\citep{Liu_2018,Liu_Zhang_Ba_Wang_Wu_2020} and inspiring the development of the interface-confined model~\citep{JAIN2024113277}. 
However, a challenge with with the diffuse interface model is its inability to accurately conserve the mass of surfactant~\citep{ERIKTEIGEN2011375,Liu_Zhang_Ba_Wang_Wu_2020}.

Three main factors contribute to the mass non-conservation of surfactant. First, the profile of the order parameter may deviate slightly from its equilibrium state due to the interplay between convection and diffusion terms~\citep{JACQMIN199996,HAO2024104750}. 
This deviation alters the profile of the regularized delta function, which subsequently affects the surfactant concentration profile and its mass conservation. 
Second, numerical diffusion, arising from discretization error of the surfactant transport equation both in space and time, results in a loss of surfactant mass~\citep{XU20125897,XU201471,CLERETDELANGAVANT2017271,ERIKTEIGEN2011375}. 
Third, in realistic long-time simulations involving the evolution of two-phase interfaces, mass conservation for each phase can not be maintained using either the level set method~\citep{XU20125897,XU201471,CLERETDELANGAVANT2017271} or the phase field method~\citep{ERIKTEIGEN2011375}.
This mass non-conservation can compromise the conservative property of the surfactant, leading to further mass loss. 


One common approach to address this issue of surfactant mass non-conservation is to rescale the surfactant concentrations at each time step to compensate for the surfactant mass loss~\citep{XU20125897,XU201471,ERIKTEIGEN2011375,Liu_Zhang_Ba_Wang_Wu_2020}.
While this technique enhances surfactant mass conservation, it can inadvertently alter the gradient of surfactant concentration.
Consequently, this alteration may hinder the accurate implementation of surface tension variation, leading to inaccuracies in simulations of two-phase flow that involve the Marangoni effect.

The primary objective of the this work is to develop an enhanced profile-preserving phase-field model for simulating two-phase flow with insoluble surfactant transport along the interface, with a focus on improving  surfactant mass conservation compared to previous models~\citep{XU20125897,ERIKTEIGEN2011375,Liu_2018,Liu_Zhang_Ba_Wang_Wu_2020,XU201471}. 
The model is developed in three main aspects: (i) We utilize the profile-preserving phase-field method from our recent work~\citep{HAO2024104750} to carefully capture the two-phase interface, preserving both the interface profiles and the regularized delta function. 
(ii) We adopt a model for the transport of interface-confined scalars~\citep{JAIN2024113277}, presenting the diffuse interface form of the surfactant transport equation within the advective Cahn-Hilliard framework.
This formulation features a gradient of the regularized delta function that is smaller than that in~\citep{ERIKTEIGEN2011375}, thereby reducing numerical diffusion. 
To mitigate time-step constraints, we employ a semi-implicit scheme to solve the surfactant transport equation and utilize the finite volume method for conservative spatial discretization.
 (iii) We introduce a new hybrid surface tension model that combines the free-energy model with the continuum-surface force model, among to minimize spatial discretization errors, particularly in two-phase flow with large density and viscosity ratios.

We organize this paper as follows. 
In Section \ref{sec:s2}, we describe the numerical model, including the Cahn-Hilliard-Navier-Stokes equations, the profile-preserving equation, the surfactant transport equation in its diffuse-interface formulation, and a hybrid surface tension model (FECSF model). 
Section \ref{sec:s3} presents the numerical methods for solving the whole governing equations in details, along with the discretization of the FECSF model. 
To demonstrate the accuracy of the numerical algorithm, Section \ref{sec:s4} provides three test cases: the evolution of surfactant concentration on an expanding drop, surfactant diffusion along the interface of a stationary spherical droplet, and spurious currents in a static drop. 
Furthermore, we showcase practical applications of our model in Section \ref{sec:s5}, considering two scenarios: the deformation of a drop containing insoluble surfactant in a simple shear flow and the oscillation of a surfactant-laden liquid filament. 
In this section, we also compare our results with those obtained from previous numerical studies. 
Finally, we conclude the paper in Section \ref{sec:s6}.


\section{Numerical Model}\label{sec:s2}

\subsection{Cahn-Hilliard-Navier–Stokes Equations and Profile-Preserving Equation}\label{sec:21CHNS}

We consider the dynamics of two immiscible, incompressible fluids separated by an interface. 
To track the motion of the interface, we employ the phase field method with a profile-preserving approach~\citep{HAO2024104750}.
This method offers improved accuracy in mass conservation for each phase and in the computation of surface tension compared to traditional phase-field methods that utilize a single interface-capturing equation, such as the Cahn-Hilliard~\citep{JACQMIN199996,DING20072078} or Allen-Cahn equations~\citep{CHIU2011185}).
This advantage is particularly significant when resolving the interface using a coarser grid.
The interface is characterized by the order parameter \(\phi(\mathbf{x},t)\), representing the phase field function, which is implicitly tracked by the advective Cahn-Hilliard equation~\citep{DING20072078}

\begin{equation} \label{EQ_CH}
    \begin{aligned}    
    &\frac{\partial \phi }{\partial t}+ \nabla \cdot \left(\phi \mathbf{u} \right) = \frac{1}{Pe_{\phi}} \nabla^2 \eta ,\\ 
    \end{aligned}
\end{equation}	
where \(\mathbf{u}\) is the dimensionless fluid velocity, \(\eta=f^{\prime}(\phi)-Cn^2 \nabla^2 \phi\) represents the dimensionless chemical potential, and \(f(\phi)=\phi^2(1-\phi)^2/4\) denotes the bulk energy density. 
The P\(\acute{\text{e}}\)clet number, defined as \(Pe_{\phi}=(M^*/Cn)^{-1}\),  characterizes the ratio of convection to diffusion for the order parameter \(\phi\), where \(M^*\) is the mobility number and \(Cn=\ell/R\) is the Cahn number, indicating the dimensionless interface thickness.
Here, \(\ell\) is the typical interface thickness at mesoscopic scales, and \(R\) is a characteristic length scale for the macroscopic size of the fluid system. 
To achieve the sharp-interface limit, where the diffuse interface converges to a sharp interface as \(Cn\) approaches zero, we employ the relations \(M^*\thicksim Cn^2\) and  \(Pe_\phi \thicksim Cn^{-1} \), as suggested by~\cite{magaletti2013sharp}. 


In a state of equilibrium, the chemical potential \(\eta\) remains constant throughout the entire domain. 
Minimizing the interfacial free energy yields the equilibrium profiles for the order parameter \(\phi(x,t)\)~\citep{Cahn1959}, expressed as a specific form is~\citep{JACQMIN199996,DING20072078}

\begin{equation} \label{EQ_EQM}
    \phi(\mathbf{x},t) =\frac{1}{2}\left\{1+\tanh \left[\frac{\psi (\mathbf{x},t)}{2\sqrt{2}Cn }\right]\right\},
\end{equation}	
where \(\psi (\mathbf{x},t)\) is the signed-distance function. 
However, the interfacial profile can iteratively deviate from this equilibrium state~\citep{JACQMIN199996} due to the interplay between convection and diffusion terms in Eq.~\eqref{EQ_CH}, leading to artificial thickening or thinning of the diffuse interface. 
To prevent the deviation, here, we use an interfacial profile-preserving approach~\citep{HAO2024104750}, which iteratively corrects the interfacial profile to maintain its equilibrium state. 
The profile-preserving equation is given by
\begin{equation} \label{EQ_PP}
       \frac{\partial \phi }{\partial \tau }=\nabla \cdot \left\{\left[\sqrt{2}Cn \left(\nabla \psi \cdot \mathbf{n(\psi)}\right)-\frac{1}{4}\left(1-{{\tanh }^{2}}\left(\frac{\psi }{2 \sqrt{2} Cn }\right)\right)\right]\mathbf{n(\psi)}\right\},
\end{equation}	
where \(\tau\) is an iteration time and \(\mathbf{n(\psi)}=\nabla \psi_{\tau=0} /\left|\nabla \psi\right|_{\tau=0}\) denotes the normal direction of the interface. 
The signed distance function \(\psi\) relates to the order parameter \(\phi\) via the expression \(\psi=\sqrt{2}Cn \ln\left(\frac{\phi}{1-\phi}\right)\). 
We iteratively solve this equation with respect to \(\tau\) until the phase field function \(\phi\) satisfies the steady state criteria, defined by
\begin{equation}\label{EQ_PP_CRITERIA}
    \int_{\Omega}\left|\phi_{m+1}-\phi_m\right| d \Omega \leq TOL \cdot \Delta \tau, 
\end{equation}
where the subscript notation m-\(th\) denotes the $m$-th artificial correction step and \(TOL\) is the threshold value. 
Unless otherwise stated, we set  \(TOL=1\) and \(\Delta \tau=\text{0.01}\Delta x\), with \(\Delta x\) representing the grid size.

The motion of two-phase flows, incorporating the effects of surface tension and gravity, is governed by the dimensionless Navier–Stokes equations, expressed as follows
\begin{equation} \label{EQ_CT}
    \ \nabla \cdot \textbf{u} = 0,\
\end{equation}
\begin{equation} \label{EQ_MOM}
    \ \rho\left(\frac{\partial \textbf{u}}{\partial t}+\mathbf{u}\cdot \nabla\mathbf{u}\right)=-\nabla p+\frac{1}{Re}\nabla \cdot \left[\mu\left(\nabla\textbf{u}+\nabla\textbf{u}^T\right)\right]+\frac{{\mathbf{F}}_{\mathbf{st}}}{We}+ \frac{\rho}{Fr}\mathbf{j},\
\end{equation}
where \(p\) represents pressure, \(\mathbf{F_{st}}\) denotes the surface tension force, and \(\mathbf{j}\) indicates the direction of gravitational acceleration. 
The properties of fluid 1 are characterized by density \(\rho_1\) and viscosity \(\mu_1\), while fluid 2  is defined by \(\rho_2\) and \(\mu_2\).
We derive the dimensionless fluid density \(\rho\) and dynamic viscosity \(\mu\) as follows,
\begin{equation}
    \begin{aligned}
        &\rho(\phi)=\phi +(1-\phi) \zeta_d , \\
        &\mu(\phi)=\phi +(1-\phi) \zeta_{\mu},       
    \end{aligned}
\end{equation}
where \(\zeta_d=\rho_2/\rho_{1}\) and \(\zeta_{\mu}= \mu_{2}/\mu_{1}\) represent the density ratio and the viscosity ratio, respectively. 
Three dimensionless parameters used in Eq.~\eqref{EQ_MOM} are Reynolds number \(Re=\rho_1 U R/\mu_1\), Weber number \(We=\rho_1 U^2R/\sigma_0\), and Froude number \(Fr=U^2/{gR}\), where \(\sigma_0\), \(g\), and \(R\) represent the surface tension coefficient, the gravitational acceleration, and the characteristic length, respectively. 
We also consider other dimensionless number used in Sections~\ref{sec:s4} and~\ref{sec:s5}, including Capillary number \(Ca=We/Re=\mu_1 U/\sigma_0\) and Ohnesorge number \(Oh=\sqrt{Ca/Re}=\mu_1/\sqrt{\rho_1 \sigma_0 R}\). 


\subsection{Surfactant Transport Equation} 

 To circumvent the calculation of the surface gradient operator \(\nabla_{s}\) and facilitate the simulations of complex topological changes, such as merging and splitting of interfaces, we employ the diffuse-interface~(DI) formulation of surfactant transport equation~(Eq.~\eqref{STE_O1})~\citep{JAIN2024113277,ERIKTEIGEN2011375}.
 In this approach, the surfactant concentration field is treated as an interface-confined scalar.

The formulation is adapted to the phase field framework alongside the advective Cahn-Hilliard equation~(Eq.~\eqref{EQ_CH}).
The relation ${\delta}\nabla \varGamma = \nabla{(\delta\varGamma)}-\varGamma \nabla {\delta}$ is employed to replace the diffusion term $\nabla \cdot \left( {{\delta }}\nabla \varGamma  \right)/Pe_{\varGamma}$ on the right-hand side of Eq.~\eqref{STE_O2}, given 
\begin{equation}\label{STE0}
    \begin{aligned}
        \frac{1}{Pe_{\varGamma}}\nabla \cdot \left( {{\delta }}\nabla \varGamma  \right)=\frac{1}{Pe_{\varGamma}}\nabla \cdot \left( \nabla{(\delta\varGamma)}-\varGamma \nabla {\delta}  \right).  
    \end{aligned}
\end{equation}
Here, the regularized delta function is defined as ${{\delta }}=\left| \nabla \phi  \right|$.
When the interfical profile is in equilibrium state, the phase field function satisfies the relation~$\nabla \phi =\phi(1-\phi)/(\sqrt{2}Cn )\nabla \psi$~\citep{WACLAWCZYK2015487}. Using the property of signed distance function $\left|\nabla \psi\right|$=1, we have 
\begin{equation}\label{STE1}
    \begin{aligned}
    \delta =\frac{\phi(1-\phi)}{\sqrt{2}Cn} \text{ and } \nabla \delta = \frac{1-2\phi}{\sqrt{2}Cn}\nabla \phi.       
    \end{aligned}
\end{equation}
We then substitute Eq.~\eqref{STE0} and Eq.~\eqref{STE1} into Eq.~\eqref{STE_O2}, and obtain
\begin{equation} \label{STE}
    \frac{\partial  {\hat{\varGamma }}}{\partial t}+\nabla \cdot \left( \hat{\varGamma }\mathbf{u} \right)= \frac{1}{Pe_{\varGamma}}  \nabla \cdot \left[ \nabla \hat{\varGamma }-\frac{(1-2\phi )\hat{\varGamma }}{\sqrt{2}Cn }\frac{\nabla \phi }{\left| \nabla \phi  \right|} \right],
\end{equation}
where $\hat{\varGamma }={{\delta }}\varGamma$.

The second term on the right-hand side of the Eq.~\eqref{STE} serves as a constraint that prevents the insoluble surfactant from diffusing out of the interface, thereby confining the diffusion within the diffuse interface region~\citep{JAIN2024113277}. 
This formulation is linked to the Cahn-Hilliard model via the parameter \(Cn\). 
Notably, the gradient of the regularized delta function used in Eq.~\eqref{STE} given by \(\delta=\phi(1-\phi)/(\sqrt{2}Cn)\) is smaller than that employed in~\citep{ERIKTEIGEN2011375}, which is \(\delta=3\sqrt{2}\phi^2(1-\phi)^2/Cn\). 
This discrepancy is notable, as numerical diffusion resulting from the  discretization of spatial derivatives~\citep{XU20125897,XU201471} cause non-conservation of surfactant mass. 
The formulation present in Eq.~\eqref{STE} within the framework of the advective Cahn-Hilliard equation, is designed to mitigate this numerical diffusion while maintaining the same discrete scheme.

It is important to emphasize that the formulation in Eq.~\eqref{STE} necessitates the interfacial profile to be in its equilibrium state.
 Consequently, the profile-preserving approach outlined in Section~\ref{sec:21CHNS} becomes crucial, as the iterative corrections to the profile enhance the mass conservation of the surfactants.

\subsection{Surface Tension Model}

As previously mentioned, the surface tension coefficient \(\sigma^\ast\) depends on the surfactant concentration \(\varGamma^\ast\), where \(\ast\) indicates dimensional variables. 
In this study, we describe the relationship between \(\sigma^\ast\) and \(\varGamma^\ast\) using the Langmuir equation of state,
\begin{equation}\label{Langmuir}
{\sigma^\ast }(\varGamma^\ast)=\sigma_0 \left[1+ \frac{\mathcal{R} T \varGamma_{\infty}^\ast}{\sigma_0} \ln \left(1-\frac{\varGamma^\ast}{\varGamma_{\infty}^\ast} \right)\right],
\end{equation}
where \(\mathcal{R}\) is the ideal gas constant, \(T\) is the temperature, and \(\varGamma_{\infty}^\ast\) is the saturation interfacial concentration of the surfactant. 
The dimensionless surface tension \(\sigma^\ast/\sigma_0\) can be expressed as 
\begin{equation}\label{n-langmuir}
\sigma (\varGamma)=1+E \ln (1-\varGamma),
\end{equation}
where \(\varGamma = \varGamma^\ast/ \varGamma_{\infty}^\ast\), and \(E  =\mathcal{R} T \varGamma_{\infty}^\ast/{{\sigma }_{0}}\) is the elasticity number. 
This number quantifies the sensitivity of surface tension changes in response to variations in surfactant concentration.

The surface tension force \(\mathbf{F_{st}}\) is compose of the capillary force \(\mathbf{F_{ca}}\)(the normal surface tension force) and the Marangoni force \(\mathbf{F_{ma}}\) (the tangential force) induced by a non-uniform distribution of surfactants. 
To compute the capillary force, we adopt a free-energy based approach as proposed in~\citep{JACQMIN199996,ding2007}. 
In the equilibrium state of the phase field \(\phi\), the capillary force associated with the surface tension coefficient \(\sigma^\ast\) is given by the  integration of the excess free energy per unit surface area across the interface~\citep{JACQMIN199996,ding2007}, resulting in 
\begin{equation} \label{EQ_ST1}   \mathbf{F_{ca}^\ast}=\frac{6\sqrt{2}\eta \nabla \phi }{Cn}\sigma^\ast. 
\end{equation}
By substituting Eq.~\eqref{n-langmuir} into Eq.~\eqref{EQ_ST1}, we derive the capillary force with a variable surface tension coefficient and its dimensionless form 
\begin{equation} \label{EQ_ST}   \frac{\mathbf{F_{ca}}}{We}=\frac{6\sqrt{2}\eta \nabla \phi }{Cn}\left[1+E \ln (1- \varGamma )\right]. 
\end{equation}
Combining this with the dimensionless Marangoni force \(\mathbf{F_{ma}}={{\nabla }_{s}}\sigma (\varGamma)\delta\) using the CSF model, we obtain the total surface tension force
\begin{equation} \label{EQ_ST2}   \frac{\mathbf{F_{st}}}{We}=\frac{6\sqrt{2}\eta\nabla \phi }{ Cn}\left[1+E \ln (1- \varGamma )\right]+{{\nabla }_{s}}\left[E \ln (1- \varGamma )\right]\delta_{st},
\end{equation}
where we choose \(\delta_{st}=\left|\nabla \phi\right|\), which is consistent with \(\delta\) used in Eq.~\eqref{STE}. 

In this model, the primary source of numerical error in calculating the capillary force stems from the spatial discretization error associated with the first-order spatial derivative of the phase-field function in Eq.~\eqref{STE}. In contrast, for the CSF model, the numerical error mainly results from the spatial discretization of the second-order spatial derivative~\(\nabla \cdot (\nabla \phi/\left|\nabla \phi\right|)\). 
Utilizing a second-order central difference scheme can effectively reduce spatial discretization errors, particularly when the interface is resolved by only a few grid cells.
This enhancement contributes to improved accuracy in the calculation of surface tension. 

We denote this hybrid Free-Energy/Continuum Surface Force model as the FECSF model, and refer to the Phase Field method utilizing this hybrid surface tension model as the PF-FECSF model.

\section{Numerical Method}\label{sec:s3}

\begin{figure}[H]
    \centering
    \includegraphics[width=0.80\textwidth]{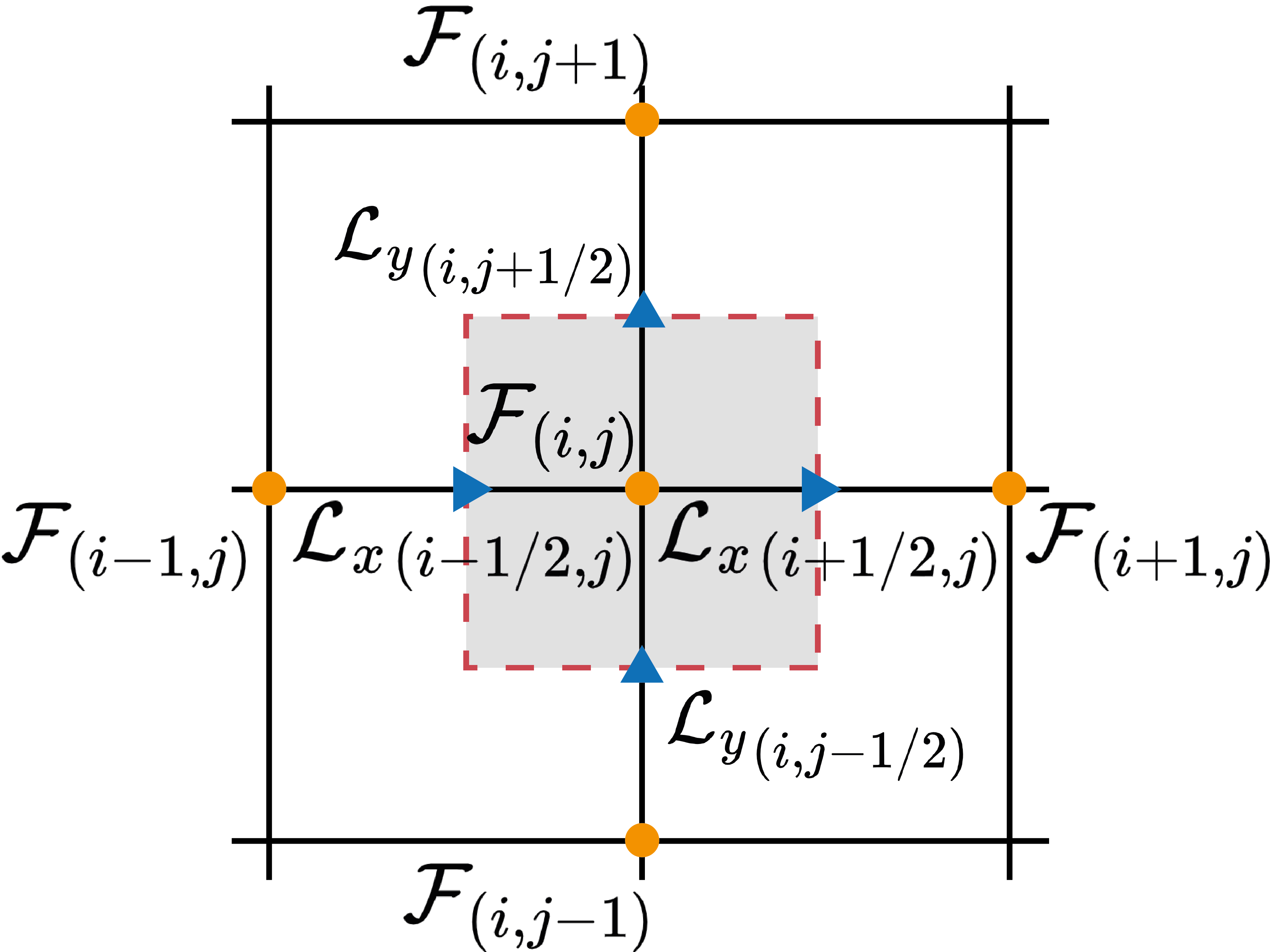}
    \caption{Configuration of the staggered mesh and the control volume.}
    \label{fig:mesh31}
\end{figure}

In this section, we describe a numerical scheme for computing the coupled equations described above. 
Specifically, we employ the finite-volume method on a rectangular uniform staggered mesh, combined with a semi-implicit time discretization.
The vector fields \(\mathbf{\mathcal{L}}=(\mathcal{L}_x, \mathcal{L}_y) \) include the velocity field \(\mathbf {u}\), the surface tension force \(\mathbf{F}_\mathbf{st}\), and the flux \(\mathbf{J}\), all defined at cell faces.
In contrast, the interface normal \(\mathbf{n}\) is defined at cell vertices, while the scalar fields \(\mathcal{F} \) = (\(p\), \(\mu\), \(\rho\), \(\phi\), \(\hat{\varGamma}\), \(\varGamma\)) are deﬁned at the cell centers, as illustrated in Fig.~\ref{fig:mesh31}.

\subsection{Temporal and Spatial Discretization}\label{sec:s31}

\begin{itemize}
    \item Adective Cahn-Hilliard equation
\end{itemize}

A semi-implicit scheme is employed to solve the advective Cahn-Hilliard equation Eq.~\eqref{EQ_CH}.
This approach combines explicit and implicit methods to effectively address the advection and nonlinear diffusion terms. 
Specifically, the advection term and the second-order diffusion term are discretized using a second-order Adams–Bashforth scheme, while the fourth-order diffusion term is treated with a Crank–Nicolson scheme to relax the time step constraint~\citep{liu2015,HAO2024104750}.
To further enhance modal damping~\citep{Ascher1995,BADALASSI2003371} and improve the stability conditions and time step,  we apply a second-order backward differentiation  formula (BDF)  to discretize \(\partial \phi/\partial t\).
Additionally, we introduce an extra linear term \(\mathcal{B}\) to the Cahn-Hilliard equation~\citep{HE2007616}, yielding the following second-order semi-implicit discretization scheme
\begin{equation}\label{TS_CH}
    \begin{aligned} 
        &\frac{{3\phi^{n+1}}-4{\phi^{n}}+{\phi^{n-1}}}{2\Delta t}=\left[\frac{3}{2}\mathcal{A} (\mathbf{u}^n, \phi^{n})-\frac{1}{2}\mathcal{A} (\phi^{n-1}, \mathbf{u}^{n-1})\right]+ \\
        &\qquad \qquad \qquad \qquad \qquad  \mathcal{B}(\phi^{n+1},\phi^{n},\phi^{n-1})-\frac{Cn^2}{2Pe_{\phi}}\left[(\nabla^4 \phi)^{n+1}+(\nabla^4 \phi)^{n}\right],
    \end{aligned}
    \end{equation}
    where \(\mathcal{A} (\phi^{n}, \mathbf{u}^n)=-\nabla \cdot \left( \mathbf{\phi \mathbf{u}} \right)^{n}+\frac{1}{Pe_{\phi}}\nabla^2{f^{\prime}(\phi)}^{n}\) and \(\mathcal{B}(\phi^{n+1},\phi^{n},\phi^{n-1})=\phi^{n+1}-2\phi^{n}+\phi^{n-1}\). 
 In this notation, the superscripts \(n\) and \(n-1\) denote the present and previous time steps, respectively, at which all quantities are known, while  the superscript~\(n+1\) indicates the subsequent time step, representing unknown quantities to be determined.

 Within the framework of the finite volume formulation, we integrate the Cahn-Hilliard equation Eq.~\eqref{EQ_CH} over a small control volume \(dV\) (represented by the gray block in Fig.~\ref{fig:mesh31}) using the divergence theorem.
 This leads to the following expression, 
    \begin{equation}\label{CH_FV}
        \begin{aligned} 
        &\int_V \frac{\partial \phi}{\partial t} dV=\int_V \left[-\nabla \cdot \left(\phi \mathbf{u}\right)+ \frac{1}{Pe_{\phi}} \left( \nabla^2{f^{\prime}(\phi)-Cn^2 \nabla^4 \phi}\right) \right]dV\\
        &\qquad \qquad =\oint_S {\left(-\mathbf{J_1}+\frac{1}{Pe_{\phi}} \mathbf{J_2}-\frac{Cn^2}{Pe_{\phi}}\mathbf{J_3}\right)} d\mathbf{s},
        \end{aligned}
    \end{equation}      
where \(\mathbf{J_1}=\phi\mathbf{u}\) and \(\mathbf{J_3} =\nabla \left(\nabla^2{\phi}\right) \). 
The flux \(\mathbf{J_2}\) can be represented as \(\mathbf{J_2}=f^{\prime\prime}(\phi) \nabla{\phi}\),  based on the relation \(\nabla \cdot \left( \nabla{f^{\prime}(\phi)}\right) =\nabla \cdot \left(f^{\prime\prime}(\phi) \nabla{\phi}\right)\).
Here, \(f^{\prime}(\phi)=\phi^3-1.5\phi^2+0.5\phi\) and \(f^{\prime\prime}(\phi)=3\phi^2-3\phi+0.5\). 
Although both formulations are mathematically equivalent, the latter form requires less smoothness of the order parameter \(\phi\) and exhibits more favorable property for discretization.
This is particularly advantageous when the order parameter deviates slightly from its equilibrium state.

To evaluate the \(\phi\) values at the cell faces for the flux \(\mathbf{J_1}\), we employ a fifth-order weighted essentially non-oscillatory (WENO) scheme~\citep{LIU1994200}, with the local flow velocity determining the upwind direction. 
For the other two fluxes, \(\mathbf{J_2}\) and \(\mathbf{J_3}\), we illustrate the 2D spatial discretization in the x direction as an example;  extending this approach to three dimensions is straightforward.
 The values of \(J_{2,(i\pm1/2,j)}\) and \(J_{3,(i\pm1/2,j)}\) at cell face, for instance, are given by
    \begin{equation}\label{242}
    \begin{aligned}
    & J_{2,(i\pm1/2,j)}= \left. \frac{f^{\prime\prime}(\phi)}{Pe_{\phi}}  \frac{\partial \phi}{\partial x} \right|  _{i \pm 1/2,j} \text{and} \quad J_{3,(i\pm1/2,j)}=\left. \frac{1}{Pe_{\phi}}  \frac{\partial}{\partial x} (\nabla^2{\phi})\right| _{i \pm 1/2,j},\\
    \end{aligned}
    \end{equation}
    where the partial derivative of \(\phi\) is approximated using standard central difference, i.e.,
    \begin{equation}\label{2435}
    \left. \frac{\partial \phi}{\partial x}\right|_{i-1/2,j}=\frac{ \phi_{i,j}-\phi_{i-1,j}}{\Delta x} \quad \text{and} \quad \left. \frac{\partial \phi}{\partial x}\right|_{i+1/2,j}=\frac{ \phi_{i+1,j}-\phi_{i,j}}{\Delta x}.
    \end{equation}

    We approximate the value of \(f^{\prime\prime}(\phi)_{i+1/2,j}\) using a second-order linear interpolation, expressed as follows,
    \begin{equation}\label{243}
        \left. f^{\prime\prime}(\phi)\right|_{i+1/2,j}=3\phi^2_{i+1/2,j}-3\phi_{i+1/2,j}+0.5, \quad \phi_{i+1/2,j}=\frac{\phi_{i+1,j}+\phi_{i,j}}{2}.
    \end{equation}
 For the spatial discretization of the Laplacian operator \(\nabla^2\), we employ standard central finite difference schemes. 
 We then solve the resulting linear system derived from the discretization of Eqs.~\eqref{TS_CH} and~\eqref{CH_FV} using the method of successive over-relaxation (SOR).
     
    \begin{itemize}
        \item Profile-preserving equation
    \end{itemize}

To discretize the profile-preserving equation Eq.~\eqref{EQ_PP} in the temporal domain, we apply the second order TVD Runge-Kutta method~\citep{gottlieb1998total}, 

\begin{equation}
    \begin{aligned}
    \phi^{(1)} & =\phi^{(0)}+\Delta \tau  \mathcal{Q} \left(\phi^{(0)}\right), \\
    \phi^{(2)} & =\frac{1}{2} \phi^{(0)}+\frac{1}{2} \Delta \tau\left[\phi^{(1)}+\mathcal{Q}\left(\phi^{(1)}\right)\right]
    \end{aligned}
    \end{equation}
    where \[\mathcal{Q}\left(\phi^{(0)}\right)=\nabla \cdot \left\{\left[\sqrt{2}Cn \left(\nabla \phi^{(0)} \cdot \mathbf{n(\phi)}\right)-\frac{1}{4}\left(1-{{\tanh }^{2}}\left(\frac{\psi }{2 \sqrt{2} Cn }\right)\right)\right]\mathbf{n(\phi)}\right\}.\] Further details regarding the spatial discretization can be found in our recent work~\citep{HAO2024104750}.
   
    \begin{itemize}
        \item Surfactant transport equation
    \end{itemize}

To circumvent time step restrictions imposed by the explicit treatment of diffusive terms~\citep{JAIN2024113277,farsoiya2024coupled}, we implement a semi-implicit scheme~\citep{Ascher1995} for the temporal discretization of the surfactant transport equation, in the same spirit of the one described in~Eq.~\eqref{TS_CH}.
 In this scheme, the advection term and sharpening terms are discretized using a second-order Adams-Bashforth scheme, while the diffusion term is discretized using a Crank-Nicolson scheme. 
 The resulting discretized surfactant transport equation is formulated as follows,
\begin{equation}\label{TS_STE}
\begin{aligned} 
    &\frac{{3\hat{\varGamma }^{n+1}}-4{\hat{\varGamma}^{n}}+{\hat{\varGamma }^{n-1}}}{2\Delta t}=-\left[\frac{3}{2}\mathcal{W} (\hat{\varGamma }^{n}, \mathbf{u}^n, \phi^{n+1})-\frac{1}{2}\mathcal{W} (\hat{\varGamma }^{n-1}, \mathbf{u}^{n-1}, \phi^{n})\right]+\\
    &\qquad \qquad \qquad \qquad \qquad  \frac{1}{2Pe_{\varGamma}}\left[(\nabla^2 \hat{\varGamma})^{n+1}+(\nabla^2 \hat{\varGamma})^{n}\right],
\end{aligned}
\end{equation}
where \[\mathcal{W} (\hat{\varGamma }^{n}, \mathbf{u}^n, \phi^{n+1})=\nabla \cdot \left( \hat{\varGamma }\mathbf{u} \right)^{n}+ \frac{1}{Pe_{\varGamma}}  \nabla \cdot \left\{\hat{\varGamma}^n \left[\frac{(1-2\phi)}{\sqrt{2}Cn }\frac{\nabla \phi }{\left| \nabla \phi  \right|}\right]^{n+1}\right\}.\]

It is worth noting that while this discretization method is not unconditionally stable~\citep{Ascher1995}, it allow for the use of larger time steps compared to explicit methods. 
To evaluate stability, we conduct numerical experiments on the diffusion of surfactant on a stationary sphere, as detailed in Section~\ref{sec:s42}.
For explicit time discretization, the constraint on the time size is given by \(\Delta t \leqslant (\Delta x)^2/2N_d D\), where \(N_d\) represents the number of dimensions. 
Our results indicate that the time step employed in the present scheme is at least an order of magnitude larger than that used in explicit time discretization. 
Furthermore, to ensure the positivity of \(\hat{\varGamma }\),  the length scale must be constrained, which can be expressed as \(\Delta x \leq 2D/( U_{max} + D/\sqrt{2}Cn)\)~\citep{JAIN2024113277}, where \( U_{max} \) is the maximum velocity magnitude at the interface.

\begin{itemize}
    \item Momentum equation
\end{itemize}

For the Momentum equation Eq.~\eqref{EQ_MOM}, we use a BDF scheme combined with the previously described semi-implicit discretization.
Specifically, the advection term is discretized using a second-order Adams–Bashforth scheme, while the viscous term is treated with a Crank–Nicolson scheme. 
To decouple the continuity equation~\eqref{EQ_CT} from  
the momentum eqution~\eqref{EQ_MOM}, we utilize a standard projection method.
The intermediate velocity \(\mathbf{u}_{im}\) is computed by
\begin{equation}\label{TS_NS}
    \begin{aligned} 
        &\rho^{n+1/2}\left(\frac{3\mathbf{u}_{im}-4\mathbf{u}^n+\mathbf{u}^{n-1}}{2\Delta t}\right)=-\rho^{n+1/2}\left[\frac{3}{2} \mathcal{H} \left(\mathbf{u}^n\right)-\frac{1}{2} \mathcal{H} \left(\mathbf{u}^{n-1}\right)\right]+\\ 
      & \qquad \qquad \frac{1}{2 Re}\left[\mathcal{L} \left(\mathbf{u}_{im}, \mu^{n+1/2}\right)+\mathcal{L} \left(\mathbf{u}^n, \mu^{n+1/2}\right)\right]+\frac{{\mathbf{F}}_{\mathbf{st}}^{n+1/2}}{We} -\frac{\rho^{n+1/2}}{Fr}\textbf{j},
    \end{aligned}
\end{equation}
where \(\mathcal{H} \left(\mathbf{u}^n\right)=(\mathbf{u}\cdot \nabla\mathbf{u})^n\) and \(\mathcal{L} \left(\mathbf{u}_{im}, \mu^{n+1/2}\right)=\nabla \cdot \left[\mu^{n+1/2}\left(\nabla\mathbf{u}_{im}\right)\right]\). Since \(\mathbf{u}_{im}\) is not divergence-free, a projection step is imposed to find the \( \mathbf{u}^{n+1}\) at time \(n+1\), as

\begin{equation}\label{pressure}
    \rho^{n+1/2} \left(\frac{3\mathbf{u}^{n+1}-3\mathbf{u}_{im}}{2\Delta t}\right)=-\nabla p^{n+1}.
\end{equation}

To ensure the divergence-free property of the velocity field, \(\nabla \cdot \mathbf{u}^{n+1}=0\) in Eq.~\eqref{EQ_CT}, we apply the divergence operator to both sides of Eq.~\eqref{pressure}.
This leads to the Poisson equation given by
\begin{equation} \label{TS_PO}
    \nabla \cdot\left(\frac{\nabla p^{n+1}}{\rho^{n+1/2}}\right)=\frac{3}{2}\frac{\nabla \cdot \mathbf{u}_{im}}{\delta t},
\end{equation}
which is solved using an efficient multigrid method with the Gauss–Siedel method as a smoother. 
Then, substituting \(p^{n+1}\) into the Eq.~\eqref{pressure}, we get the velocity \(\mathbf{u}^{n+1}\).

\subsection{Discretization of Surface Tension Force Formulation}

We present the discretization of the surface tension force as described in Eq.~\eqref{EQ_ST2}, focusing on a two-dimensional spatial framework, which can be readily extended to three-dimensions. 
The surface tension force is expressed as the sum of the capillary force and the Marangoni force, and its spatial discretization can be approached as follows.

The capillary force, $ \mathbf{F_{ca}}=\left(F_{ca,x}, F_{ca,y}\right)$, where 
\begin{equation}\label{TS_fecsf_ca}
    \begin{aligned}
    & \left.F_{ca,x}\right|_{i+1/2,j}=\left. \frac{6\sqrt{2}\eta \left[1+E \ln (1- \varGamma )\right] }{Cn}\frac{\partial \phi}{\partial x}\right|_{i+1/2,j},\\
    & \left.F_{ca,y}\right|_{i,j+1/2}=\left. \frac{6\sqrt{2}\eta \left[1+E \ln (1- \varGamma )\right] }{Cn}\frac{\partial \phi}{\partial y}\right|_{i,j+1/2},\\
    & 
    \end{aligned}
    \end{equation}
    and the Marangoni force, $\mathbf{F_{ma}}=\left(F_{ma,x}, F_{ma,y}\right)$, where
    \begin{equation}\label{TS_fecsf_ma}
        \begin{aligned}
        & \left.F_{ma,x}\right|_{i+1/2,j}=\left. \left[\left(1-n_x^2\right)\frac{\partial \sigma(\varGamma)}{\partial x}-n_xn_y\frac{\partial \sigma(\varGamma)}{\partial y}\right] \sqrt{\left(\frac{\partial \phi}{\partial x}\right)^2+\left(\frac{\partial \phi}{\partial y}\right)^2} \right|_{i+1/2,j},\\
        & \left.F_{ma,y}\right|_{i,j+1/2}=\left. \left[-n_xn_y \frac{\partial \sigma(\varGamma)}{\partial x} + \left(1-n_y^2\right)\frac{\partial \sigma(\varGamma)}{\partial y} \right] \sqrt{\left(\frac{\partial \phi}{\partial x}\right)^2+\left(\frac{\partial \phi}{\partial y}\right)^2} \right|_{i,j+1/2}.\\
        & 
        \end{aligned}
        \end{equation}
\(n_x\) and \(n_y\) are the components of the normal vector \(\mathbf{n}\) in the \(x\) and \(y\) directions, respectively, written as 
\begin{equation}\label{TS_fecsf_n}
    \begin{aligned}
    &\left. n_x\right|  _{i+1/2,j+1/2}=\left. \frac{\frac{\partial \phi}{\partial x}}{\sqrt{\left(\frac{\partial \phi}{\partial x}\right)^2+\left(\frac{\partial \phi}{\partial y}\right)^2}}\right|_{i+1/2,j+1/2},\\
    &\left. n_y\right|_{i+1/2,j+1/2}=\left. \frac{\frac{\partial \phi}{\partial y}}{\sqrt{\left(\frac{\partial \phi}{\partial x}\right)^2+\left(\frac{\partial \phi}{\partial y}\right)^2}}\right|_{i+1/2,j+1/2}.\\
    &
    \end{aligned}
\end{equation}

In Eqs.~\eqref{TS_fecsf_ca}-\eqref{TS_fecsf_n}, we utilize a second-order central difference scheme to approximate the gradient operator, while a second-order linear interpolation scheme is applied to approximate the values of variable at cell face (\(i \pm 1/2,j\) or \(i,j \pm 1/2\)). 
For example, if \(\varGamma\) and \(\eta\) are solved at \((i+1/2,j)\), then
        \begin{equation}\label{cell_face}
        \begin{aligned}
        & {\varGamma}_{i+1/2,j} = \frac{1}{2}({{\varGamma}_{i+1,j}}+{{\varGamma}_{i,j}}),\\
        & {\eta}_{i+1/2,j} = \frac{1}{2}({{\eta}_{i+1,j}}+{{\eta}_{i,j}}).\\
        &
        \end{aligned}
        \end{equation}

\subsection{Solution Procedure}

\begin{figure}[H]
    \centering
    \includegraphics[width=0.6\textwidth]{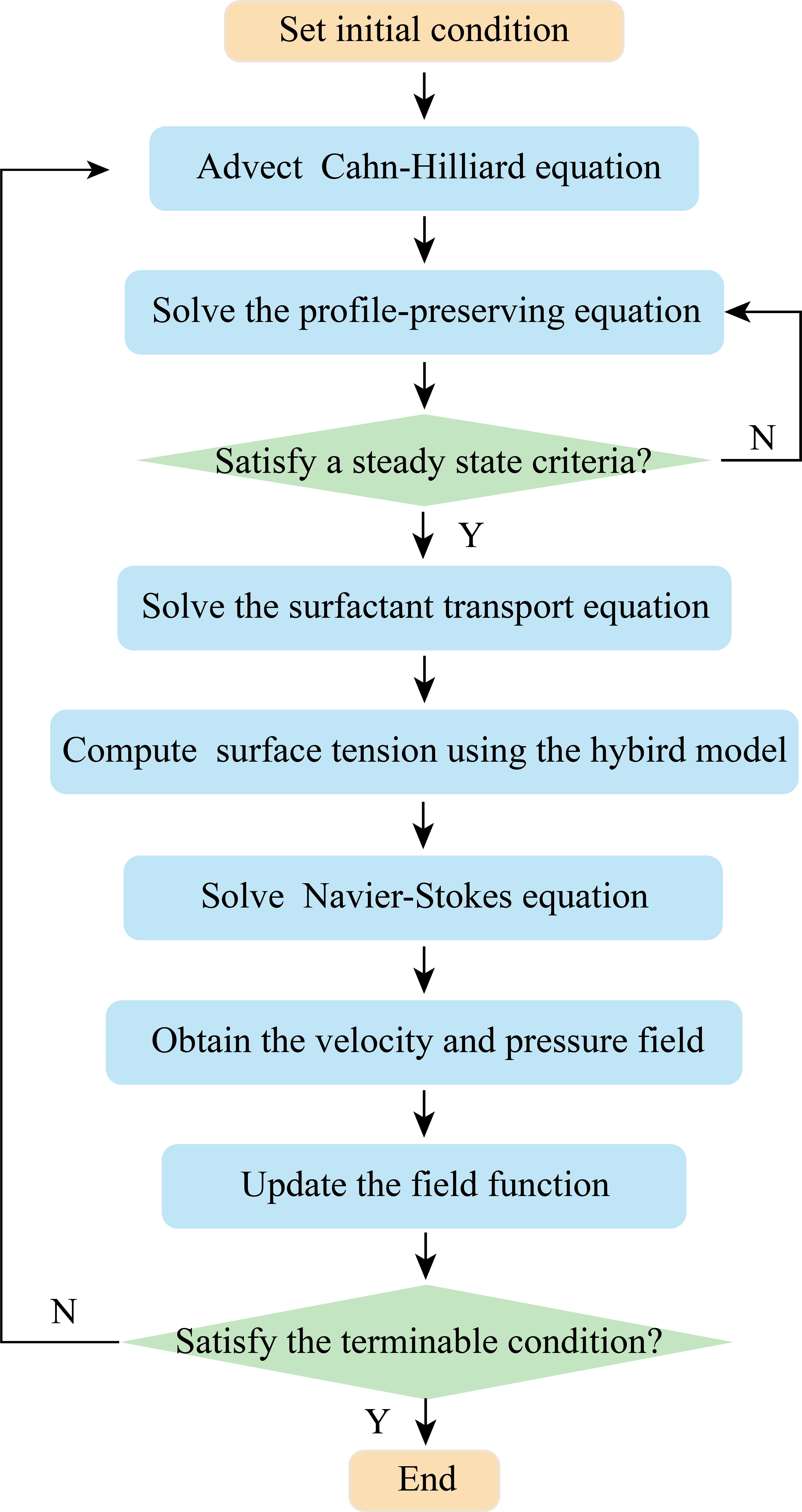}
    \caption{Flow chart of the computational procedure.}
    \label{fig:mesh32}
\end{figure}

As depicted in Fig.~\ref{fig:mesh32}, the comprehensive solution procedure for each time step loop can be summarized as follows

\begin{enumerate}
\item[(1)] Advance the phase field function \(\phi(\mathbf{x},t)\) using Eq.~\eqref{TS_CH} after initialization. 
\item[(2)] Correct the profile of the phase field \(\phi(\mathbf{x},t)\) using the profile-preserving equation Eq.~\eqref{EQ_PP} and update the phase field function \(\phi(\mathbf{x},t)\).
\item[(3)] Solve the surfactant transport equation using Eq.~\eqref{TS_STE} and compute the surfactant concentration \(\varGamma(\mathbf{x},t)\).
\item[(4)] Compute the capillary force and Marangoni force using Eq.~\eqref{EQ_ST}.
\item[(5)] Compute the intermediate velocity~\(\mathbf{u}^*\) by solving Eq.~\eqref{TS_NS}.
\item[(6)] Make the velocity divergence using the continuity equation Eq.~\eqref{EQ_CT} and solve the Poisson equation Eq.~\eqref{TS_PO} using the multigrid method. 
\item[(7)] Obtain the new pressure field~\(p^{n+1}\) and the velocity~\(\mathbf{u}^{n+1}\), and update these functions.

\end{enumerate}   

\section{Numerical Validation}\label{sec:s4}

In this section, we first perform two standard numerical tests proposed by~\cite{MURADOGLU20082238} to assess the accuracy of the numerical solution algorithm. 
The first test evaluates the mass conservation of surfactant at the interface of an expanding circle, as well as the mesh convergence rate of the algorithm. 
The second test examines the diffusion of surfactant on a stationary spherical drop, where we compare the numerical results with analytical solutions. 
Finally, we simulate a spurious current flow in a stationary surfactant-laden drop to evaluate the performance of the FECSF model, and we compare these results with those obtained using the CSF model.

\subsection{Surfactant Concentration Evolution on an Expanding Circle}\label{sec:s41}

To evaluate the accuracy of the advection term \(\nabla \cdot ( \hat{\varGamma }\mathbf{u})\) in surfactant transport equation (Eq.~\eqref{STE}) and its temporal and spatial discretization (Eq.~\eqref{TS_STE}), we perform simulations of the evolution of surfactant concentration on an expanding drop in a two dimensional space, as proposed by~\cite{MURADOGLU20082238}. 
In this test, we assume the initial surfactant concentration \({\varGamma}_{0}\) is uniform, and the radial velocity field is given by \(\mathbf{u}=(\cos \theta, \sin \theta)\).
Under these conditions, the surfactant concentration \(\varGamma(t)\) evolves according to \(\varGamma \left( {{\nabla }_{s}}\cdot \mathbf{n} \right)(\mathbf{u}\cdot \mathbf{n})\), as described by Eq.~\eqref{STE_O1}. 
By integrating this equation over time and two-dimensional space, we derive the following surfactant mass conservation relationship,
\begin{equation} \label{411}
    \varGamma(t) =\frac{R_0}{R(t)}{\varGamma}_{0},
\end{equation}
where \(R_0=0.5\) is the initial radius of drop, and the instantaneous radius $R(t)$ evolves as \(R(t)=R_0+\left\lvert\mathbf{u}\right\rvert  t\).
This equation demonstrates that as the drop expands, the surfactant concentration decreases inversely with the increasing radius, while maintaining the total surfactant mass on the interface.

\begin{figure}[H]
    \centering
    \includegraphics[width=0.99\textwidth]{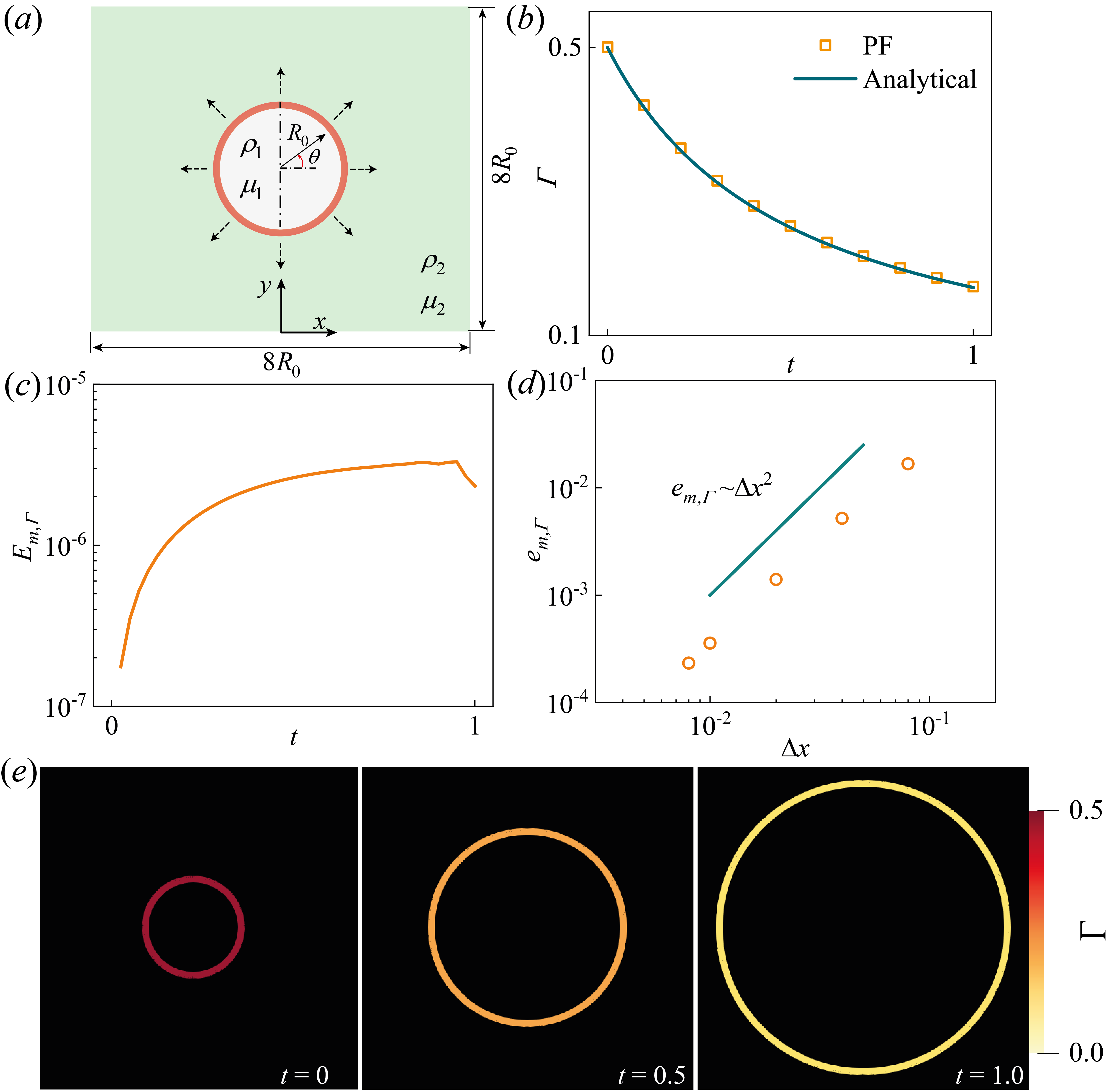}
    \caption{Advection of surfactant concentration as a circular interface expands. (a) Schematic representation of the surfactant concentration evolution on an expanding interface. (b) Comparison of temporal evolution of surfactant concentration \(\varGamma \) with analytical results at \(R_0/\Delta x=50\) and \(Cn=0.75\Delta x\). (c) Surfactant mass error \(E_{m, \varGamma}(t)\) as a function of time, demonstrating the conservation accuracy. (d) Convergence analysis of the surfactant concentration error \(e_{m, \varGamma}(t)\) at \(t = 1\), illustrating second-order accuracy. (e) Contours of the interface \(\phi=0.5\), colored by surfactant concentration \(\varGamma \), at different time instances \(t=0, 0.5 \), and \(1\).}
    \label{fig:mesh412}
\end{figure}

Unlike previous works~\citep{MURADOGLU20082238,STRICKER2017467,JAIN2024113277,farsoiya2024coupled}, we compute the two terms \(\frac{1}{Pe_{\varGamma}}  \nabla \cdot \left[ \nabla \hat{\varGamma }-\frac{(1-2\phi )\hat{\varGamma }}{\sqrt{2}Cn }\frac{\nabla \phi }{\left| \nabla \phi  \right|} \right]\) on the right-hand side of Eq.~\eqref{STE}.
This enables us to not only evaluate the performance of semi-implicit discretization for the Eq.~\eqref{TS_STE}, but also to assess the mesh convergence of our method.
 As shown in Fig.~\ref{fig:mesh412} (a), the computations are performed in a square domain of size \(8R_0 \times 8R_0\), with a circle initially placed at the center (\(4R_0, 4R_0\)) and undergoing continuous, uniform expansion.
 The dimensionless parameters used in these simulations are: initial surfactant concentration \({\varGamma}_{0}=0.5\), the \(Pe_{\phi}\) number  for the phase field \(Pe_{\phi}=1/Cn\), and the \(Pe_{\varGamma}\) number for the surfactant \(Pe_{\varGamma}=1\).

The surfactant concentration \(\varGamma\) decreases due to the normal expansion of the circular interface, as shown in Fig.~\ref{fig:mesh412} (b), which aligns well with the analytical solution provided by the Eq.~\eqref{411}.
The surfactant concentration remains uniformly distributed along the interface, as illustrated in Fig.~\ref{fig:mesh412} (e).
To evaluate the performance of surfactant mass conservation, we define the surfactant mass error \(E_{m, \varGamma}(t)\) as \(E_{m,\varGamma}(t) = \int_{S }{\left| {{\varGamma(t) }}-{{\varGamma}_{0}} \right|}dS / \int_{S} {{\varGamma}_{0}} dS= \int_{\Omega }{| {{\hat{\varGamma}(t) }}-{{\hat{\varGamma} }_{0}} |}dV / \int_{\Omega }{{\hat{\varGamma} }_{0}}dV\) based on the Eq.~\eqref{STE_a2}. 
The numerical results show that $E_{m,\varGamma}(t)$ remains close to $10^{-6}$, as shown in Fig.~\ref{fig:mesh412} (c), indicating good conservation of surfactant mass. 

To assess the mesh convergence of the algorithm, we define the surfactant concentration error $e_{m, \varGamma}(t)$ between numerical and analytical results as $e_{m, \varGamma}(t) = \left| {{\varGamma(t) }}-{\frac{R_0}{R_0+\left\lvert\mathbf{u}\right\rvert t}{\varGamma}_{0}} \right| / {\frac{R_0}{R_0+\left\lvert\mathbf{u}\right\rvert t}{\varGamma}_{0}} $. In Fig.~\ref{fig:mesh412} (d), we present the convergence rate of $e_{m, \varGamma}(t)$ at $t=1$ for a fixed numerical interface thickness $\epsilon _t=0.125R_0$, which is defined in the range $0.05\leq \phi(x)\leq 0.95$~\citep{JACQMIN199996,HAO2024104750}. 
The results indicate a convergence rate of approximately 2, confirming second-order spatial accuracy of the algorithm.

\subsection{Surface Diffusion on a Stationary Sphere}\label{sec:s42}

\begin{figure}[H]
    \centering
    \includegraphics[width=1.0\textwidth]{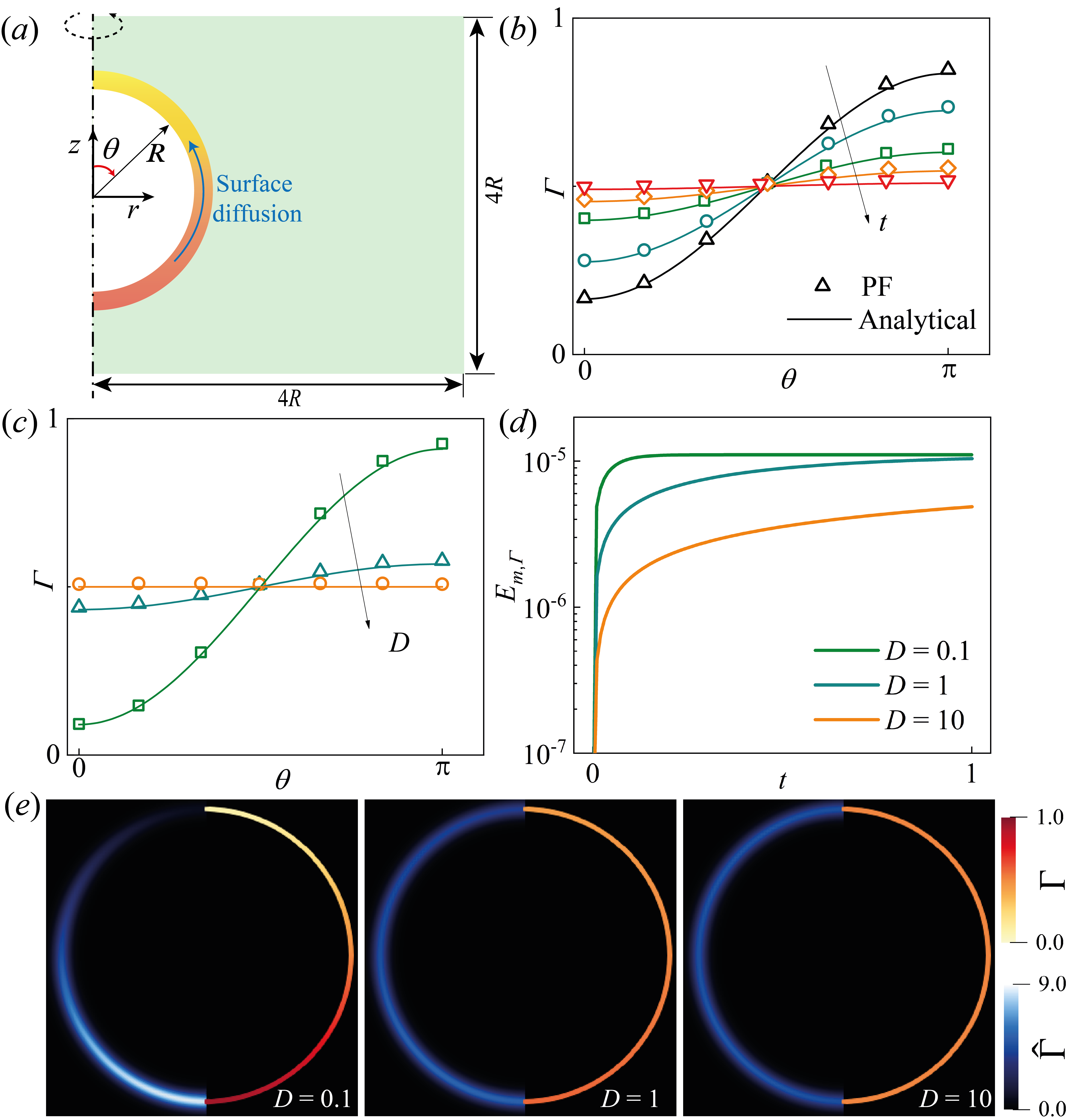}
    \caption{Diffusion of surfactant concentration at a spherical interface. (a) Schematic representation of surfactant diffusion on a spherical interface. (b) Temporal evolution of \(\varGamma \) (dots) as a function of the polar angle \(\theta\), compared with analytical results (solid lines) at \(t=0.2,0.4,0.8,1.2\), and 2.0 for the diffusion coefficient \(D=1\). 
    (c) Comparison of \(\varGamma \) distribution (dots) as a function of \(\theta\) with analytical results (solid lines) at \(t=1\) for diffusion coefficients \(D=0.1, 1\) and \(10\). 
    (d) Surfactant mass error \(E_{m, \varGamma}(t)\) as a function of time. (e) numerical snapshots of the surfactant concentration distributions at the interface, with $\hat{\Gamma}$ shown on left side and $\Gamma$ on the right side, at \(t=1\) for \(D=0.1, 1\) and \(10\).}
    \label{fig:mesh422}
\end{figure}

To further evaluate the accuracy of the semi-implicit temporal discretization and the spatial discretization for the Eq.~\eqref{TS_STE}, we simulate surfactant diffusion on a stationary spherical interface~\citep{MURADOGLU20082238,Muradoglu2014,Shin2018,JAIN2024113277}. 
Fig.~\ref{fig:mesh422} (a) illustrates the schematic of the computational domain, which measures \(4R \times 4R\), with a sphere of radius \(R\) positioned at \(z=2R\). 
The Capillary number and mesh size are defined as \(Cn=0.75\Delta x\) and \(R/\Delta x=50\). 
The time step \(\Delta t\) is set to $10^{-4}$ for \(D=10\), while it is adjusted to $10^{-3}$ for \(D=0.1\) and \(D=1\). 
An axisymmetric boundary is implemented at \(r=0\), and the Neumann boundary conditions are applied at the remaining boundaries.

When the advection term in Eq.~\eqref{STE_O1} is neglected, the surfactant transport equation simplifies to the diffusion equation on a spherical interface, expressed as
\begin{equation} \label{421}
    \ \frac{\partial \varGamma }{\partial t} = D\nabla _{s}^{2}\varGamma.\ 
\end{equation}
Assuming an initial surfactant concentration by \(\varGamma_0(\theta) =0.5(1-\cos \theta )\), as illustrated in Fig.~\ref{fig:mesh422} (a), the analytical solution for \(\varGamma (\theta ,t)\) can be expressed~\citep{MURADOGLU20082238}
\begin{equation} \label{422}
    \varGamma (\theta ,t)=0.5(1-{{e}^{-2Dt/{{R}^{2}}}}\cos \theta ).
\end{equation}
As shown in Fig.~\ref{fig:mesh422} (b), the calculated surfactant concentration distribution \(\varGamma\) from the simulation with $D=1$, is represented by dots and follows the analytical solution at various time instances ($t= 0.2, 0.4 ,0.8, 1.2$, and $2$). 
Surfactants diffuse along the interface due to concentration gradient, resulting in \(\varGamma \approx 0.5\) at the final time of \(t=2\). 

For various surface diffusion coefficients, the calculated surfactant concentration distribution \(\varGamma\) becomes increasingly flattened with rising diffusion coefficients.
Fig.~\ref{fig:mesh422} (c) presents the surfactant concentration distribution curves at the time instance \(t=1\) for \(D=0.1, 1\), and 10 (represented by dots), all aligning closely with the corresponding analytical solution (solid lines).
Fig.~\ref{fig:mesh422} (e) depicts numerical snapshots of the surfactant concentration distributions at the interface, with $\hat{\Gamma}$ shown on left side and $\Gamma$ on the right side.
Additionally, the defined surfactant mass error \(E_{m, \varGamma}(t)\) was calculated and is displayed in Fig.~\ref{fig:mesh422} (d).
All the maximum values for different diffusion coefficients are less than  \(10^{-5}\), indicating that the numerical model demonstrates good performance in surfactant mass conservation.

\subsection{Spurious Currents in a Stationary Drop}\label{sec:s43}

The accuracy of simulations involving urface-tension-driven flows is critically influenced by the surface tension models and its implementation.
In this context, the discrete approximation at the interface can introduce  imbalance between the surface tension force and the Laplace pressure difference, resulting in the generation of artificial flows, commonly referred to as spurious currents~\citep{Popinet2018}. 
The intensity of these spurious currents depends on the performance of the employed surface tension model~\citep{ABADIE2015611} and the specific  numerical discretization scheme used~\citep{Popinet2018}. 
Therefore, evaluating the spurious currents is essential for accurately simulating interfacial phenomena, particularly those involving Marangoni effects, such as the self-propulsion of droplets driven by concentration gradients of solute~\citep{STRICKER2017467,lxxb2024-011}. 



To numerically evaluate the performance of the FECSF model (Eq.~\eqref{EQ_ST2}), we conducted an analysis of spurious currents in a static drop, comparing the results with those obtained from the CSF model. Previous studies indicate that the density ratio has an impact on the intensity of spurious current for the CSF model~\citep{ABADIE2015611}. 
In this test, we extend the analysis to a more general two-phase flow scenario by varying both the viscosity ratio and the density ratio.
This allows us to assess their combined effects on the intensity of spurious currents for both the FECSF and CSF models.




We conduct two-dimensional simulations within a domain of \(2R \times 4R\), positioning a surfactant-laden drop at the center, as illustrated in Fig.~\ref{fig:mesh422} (a). 
All boundaries are subjected to no-slip boundary conditions, except for the left boundary, where a symmetry boundary condition is applied.
We assume that the spurious currents are balanced by the inertial term; therefore, we choose the drop radius \(R\) as the characteristic length and the inertial-capillary time $\sqrt{\rho R^3/\sigma}$ as the characteristic time. 
This leads to a \(We\) number of \(We=1\) in Eq.~\eqref{EQ_MOM}. 
We set \(Re=10\), \(R/\Delta x=64\), and \(\Delta t=2.5\times10^{-4}\). Additionally, we initialize the surfactant concentration at the drop interface with a uniform value of  \(\Gamma_0=0.5\) and activate the Marangoni force.
Although the Marangoni force is theoretically zero under these conditions, discretization error still arises in the implementation of the Marangoni force.

\begin{figure}[H]
    \centering
    \includegraphics[width=1.0\textwidth]{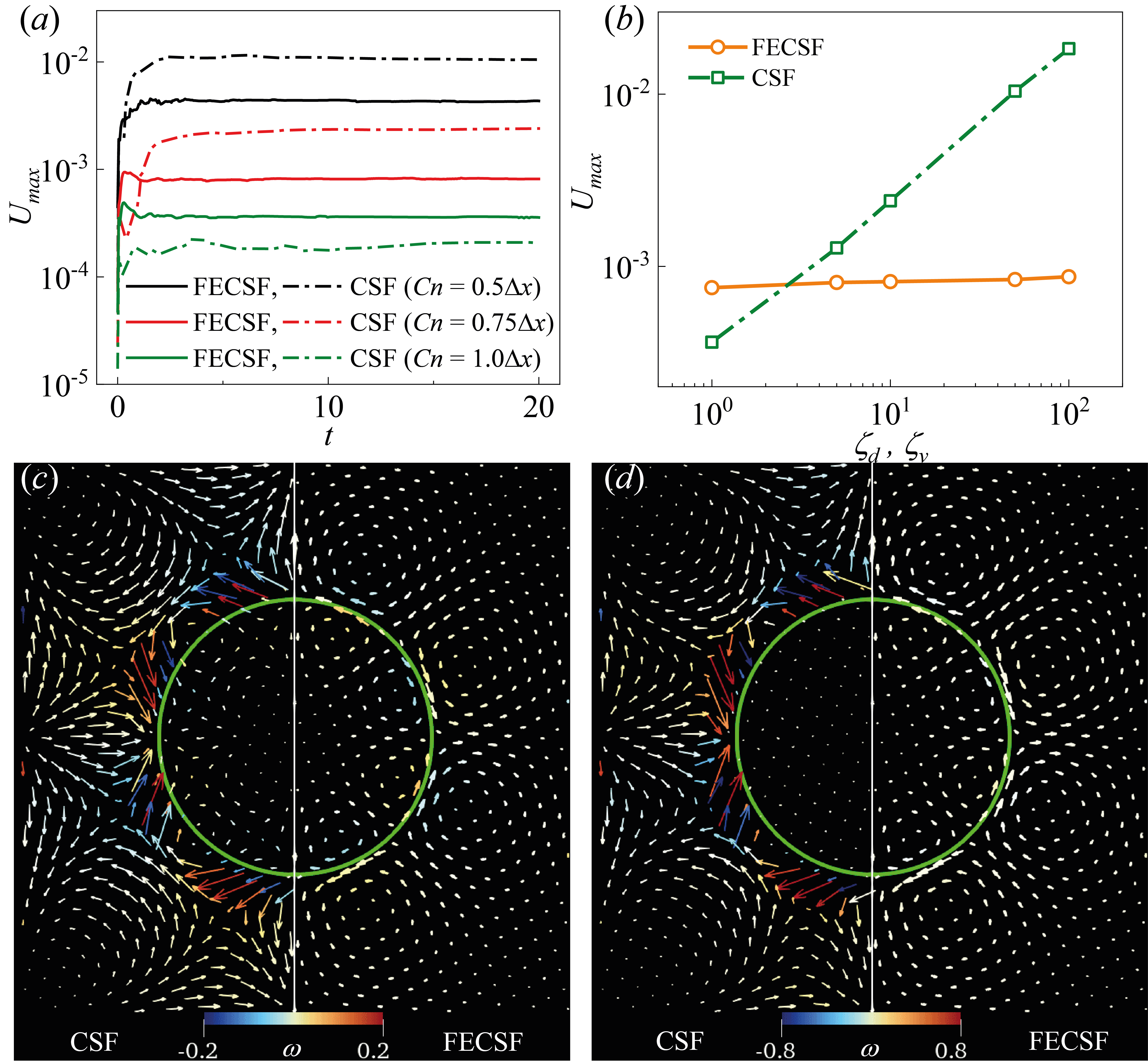}
    \caption{Spurious currents around for a static drop with an initially uniform surfactant concentration. (a) Temporal evolution of the maximum dimensionless velocity magnitude \(U_{max}\) for the density ratio \(\zeta_d =10\) and the viscosity ratio \(\zeta_{\mu} =10\). 
    (b) Maximum steady-state velocity magnitude \(U_{max}\) at various density and viscosity ratios for \(Cn=0.75\Delta x\). 
    (c) and (d) display the steady-state vorticity fields (color-coded) and  velocity vectors (arrows) for the CSF model (left) and FECSF model (right), with \(\zeta_d = \zeta_{\mu} = 10\) in (c) and \(\zeta_d = \zeta_{\mu} = 100\) in (d). 
    In (d), the length of arrows in the left panel has been reduced by a factor of four to better visualize the velocity field pattern.
} 
    \label{fig:mesh432}
\end{figure}

We compute the maximum velocity magnitude \(U_{max}\) in the domain to evaluate the intensity of spurious currents. 
Fig.~\ref{fig:mesh432} (a) illustrates the spurious current flow for different interface thicknesses \(\epsilon_t\) of the drop (\(Cn = 0.5, 0.75\), and 1.0\(\Delta x\)) at viscosity and density ratios \(\zeta_d = \zeta_{\mu} =10\). 
The results show that the FECSF model produces lower spurious currents compared to the CSF model for \(Cn= 0.5\) and 0.75. 
However, when \(Cn= 1.0\), where a greater number of grid points resolve the  interface, the CSF model demonstrates lower spurious currents.
This improvement is likely due to the reduction of discretization errors in the curvature estimation. 



We investigate the impact of the viscosity ratio \(\zeta_{\mu}\) and the density ratio \(\zeta_d\) on the intensity of spurious currents at \(Cn=0.75\Delta x\). 
Fig.~\ref{fig:mesh432} (b) demonstrates that the spurious current intensity remains minimal in the FECSF model, even as \(\zeta_{\mu}\) and \(\zeta_d\) increase.
In contrast, the CSF model shows a significant rise in spurious current intensity under the same conditions. 
The peak velocity magnitude is observed outside the drop, where the fluid has lower density and viscosity, as shown in Figs.~\ref{fig:mesh432} (c) and (d). 
These results indicate that the FECSF model provides a more balanced and stable performance compared to the CSF model, especially in simulation involving large density and viscosity ratios in two-phase flows.

\section{Applications}\label{sec:s5}

To demonstrate the applicability of the numerical algorithm, 
we first simulate the deformation of a droplet with insoluble surfactant in a typical shear flow. 
This setup is commonly employed to regulate the dynamics of emulsion systems containing surfactants~\citep{Nandi1999,Cates_Tjhung_2018}. 
Following this, we study the contraction and oscillation dynamics of a filament in the presence of surfactant, a critical process in the drop-on-demand (DOD) inkjet system~\citep{Manikantan2020,detlef2022,Wang_Zhong_Fang_2019}. The evolution of surfactant concentration distribution is presented, and the effects of capillary and Marangoni forces on the oscillation amplitude and period are discussed.



\subsection{Deformation of Drop with Insoluble Surfactant in a Simple Shear Flow}\label{sec:s51}

\begin{figure}[H]
    \centering
    \includegraphics[width=1.0\textwidth]{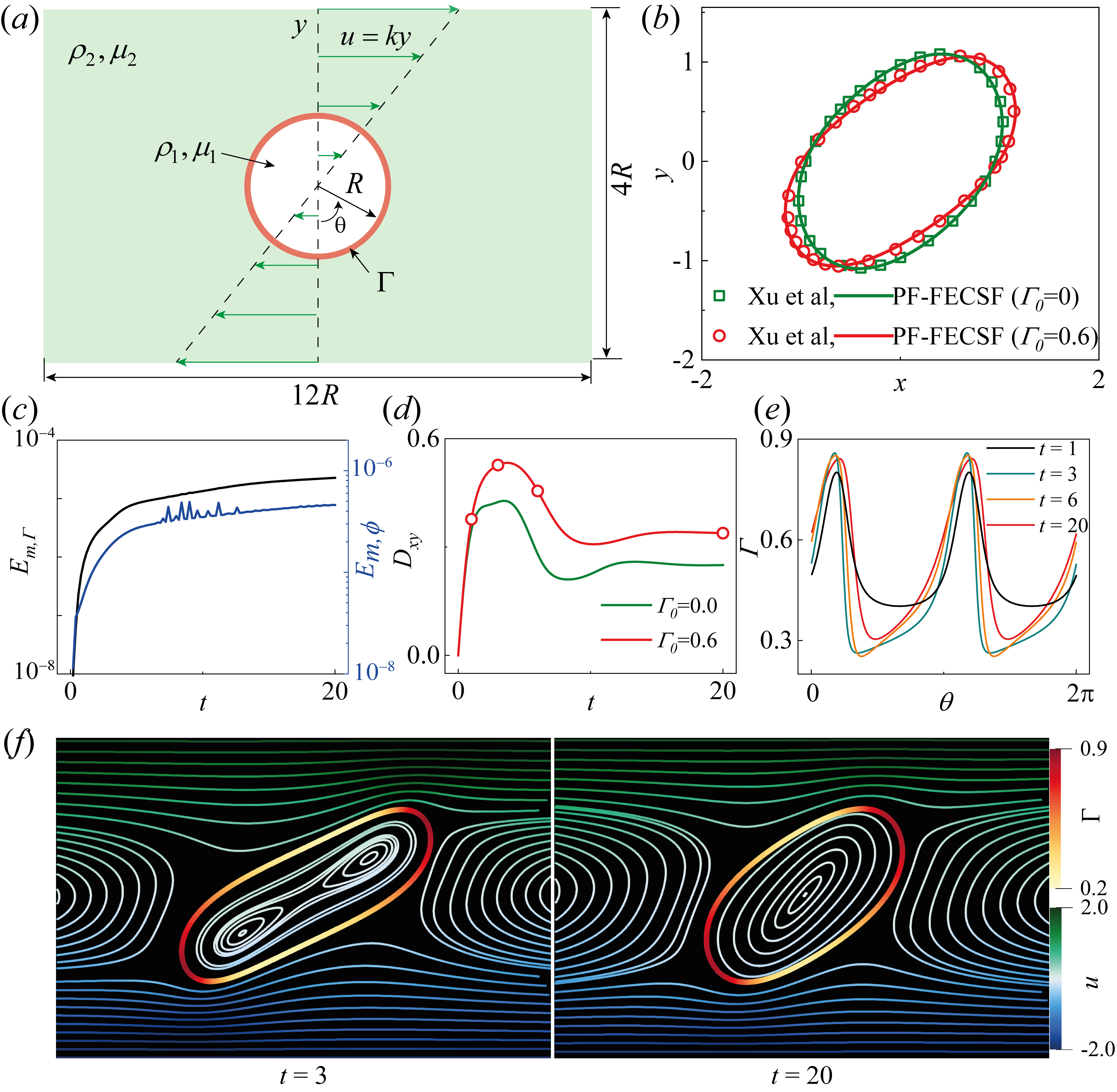}
    \caption{Deformation of a 2D drop with and without insoluble surfactant in a simple shear flow. 
    (a) Schematic of a drop with insoluble surfactant at its interface in a simple shear flow. 
    (b) Comparison of drop deformation at \(t=9\) for surfactant-free (\(\varGamma_0=0\))  and surfactant-covered (\(\varGamma_0=0.6\))  drops, alongside reference results from the level set method~\citep{XU20125897}. (c) Temporal evolution of mass error for surfactant concentration, \(E_{m,\varGamma}\), and fro the drop phase, \(E_{m, \phi}\). 
    (d) Temporal evolution of the Taylor deformation parameter \(D_{xy}\) for both  \(\varGamma_0=0\) and \(\varGamma_0=0.6\). 
    (e) Distribution of surfactant concentration \(\varGamma \) as a function of polar angle \(\theta\) at \(t=1, 3, 6\), and 20 (marked in red circles in (d)) for \(\varGamma_0=0.6\). 
    (f) Drop shape, surfactant distribution along the interface, and streamline patterns at \(t=3\)~(left) and \(t=20\)~(right).}
    \label{fig:mesh512}
\end{figure}

We investigate the deformation of a 2D drop with and without surfactant in a simple shear flow to show the applicability and accuracy of the PF-FECSF method. 
As illustrated in Fig.~\ref{fig:mesh512} (a), a drop with radius \(R\) and a uniform initial surfactant concentration distribution \(\varGamma_0\) is initially placed at the center of the computational domain. 
The simulations are conducted on a domain of size \(12R \times 4R\) under a simple shear flow with velocity \(\mathbf{u}=(ky,0)\), where \(k\) is the shear rate. 
The characteristic length and time scales are defined as \(R\) and \(1/k\), respectively, while the characteristic velocity, Reynolds number \(Re\), and Capillary number \(Ca\) are given by \(kR\), \(Re=\rho_1 k R^2/\mu_1\), and \(Ca= \mu_1 k R/\sigma\), respectively.

The drop deformation is quantified using the Taylor deformation parameter \(D_{xy}\), defined as \(D_{xy}=(R_L-R_B)/(R_L-R_B)\),  where \(R_L\) and \(R_B\) denotes the drop's longest and shortest axes. 
A no-slip boundary condition is applied in the \(y\) direction, while periodic boundary conditions are imposed in the \(x\) direction.
To compare our results with those from~\citep{XU20125897}, we choose dimensionless parameter as follows: \(Re=10\), \(Ca=0.1\), \(Cn=0.75\Delta x\), \(Pe_{\phi}=0.9/Cn\), \(E=0.2\), and \(Pe_{\varGamma}=10\). 
The mesh size, time step, the density ratios, and viscosity ratio are set to \(R/\Delta x=64\), \(\Delta t = 5\times 10^{-4}\), \(\rho_1/\rho_2=1\), and \(\mu_1/\mu_2=1\), respectively.




Fig.~\ref{fig:mesh512} (b) shows the instantaneous shape of the drop at \(t=9\) for both the clean case (\(\varGamma_0=0\)) and the surfactant-laden case (\(\varGamma_0=0.6\)). 
Our results exhibit strong agreement with those obtained using the level set method~\citep{XU20125897}, validating the accuracy and reliability of the FECSF model. 
Fig.~\ref{fig:mesh512} (c) gives the mass error of surfactant concentration \(E_{m,\varGamma}\) and drop \(E_{m, \phi}\) as functions of time, where \(E_{m, \phi}\) is defined as \(E_{m, \phi}= \int_{\Omega }{\left| {{\phi }}-{{\phi }_{0}} \right|}dV / \int_{\Omega }{\phi }_{0}dV\). 
The total surfactant mass remains nearly conserved during the drop's deformation, with the mass error remaining below \(0.002\%\). 
Moreover, the surfactant mass error \(E_{m,\varGamma}\) is reduced by nearly two orders of magnitude compared to the results obtained using the diffuse interface method~\citep{ERIKTEIGEN2011375}. 
Thus, our model eliminates the need for a mass-correction scheme, unlike the approaches in~\citep{ERIKTEIGEN2011375,XU20125897,Liu_2018}, where the surfactant concentration is adjusted by a factor to preserve the mass.

In the case of a clean drop, it undergoes elongation in response to shear stress exerted by the surrounding fluid. This elongation continues until a force balance is achieved between the surface tension and the shear stress~\citep{Sibillo2006}. This dynamics depends on the \(Re\) number and the \(Ca\) number. For a surfactant-covered drop, its deformation increases since the presence of the surfactant leads to a reduction in the surface tension~\citep{XU20125897,ERIKTEIGEN2011375,Liu_2018}.

The presence of surfactant increases drop deformation by reducing surface tension, as depicted in Fig.~\ref{fig:mesh512} (b), with a quantitative comparison in Fig.~\ref{fig:mesh512} (d) showing the temporal evolution of the Taylor deformation parameter $D_{xy}$. 
Both surfactant-laden and surfactant-free drops display damped oscillations driven by inertia, which is evident in the numerical snapshots at $t=3$ and $t=20$, shown in Fig.~\ref{fig:mesh512} (f).
The surfactant concentration distribution along the drop interface during deformation is presented in Fig.~\ref{fig:mesh512} (e). 
Interestingly, this distribution exhibits non-monotonic behavior over time. 
In the stretch stage, the shear flow stretches the drop and surfactant accumulates at the tips of the drop due to convection at \(t=3\). After that, the drop begins to contract, leading to a decrease in the curvature of interface at tips of drop in Fig.~\ref{fig:mesh512} (f) when the drop reaches equilibrium at \(t=20\). This means that a normal dilatation occurs in drop tips, similar to Fig.~\ref{fig:mesh412}, causing the surfactant concentration to diminish at \(t=20\).








\subsection{Contraction and Oscillation of Surfactant-Laden Liquid Filament}\label{sec:s52}

\begin{figure}[H]
    \centering
    \includegraphics[width=1.0\textwidth]{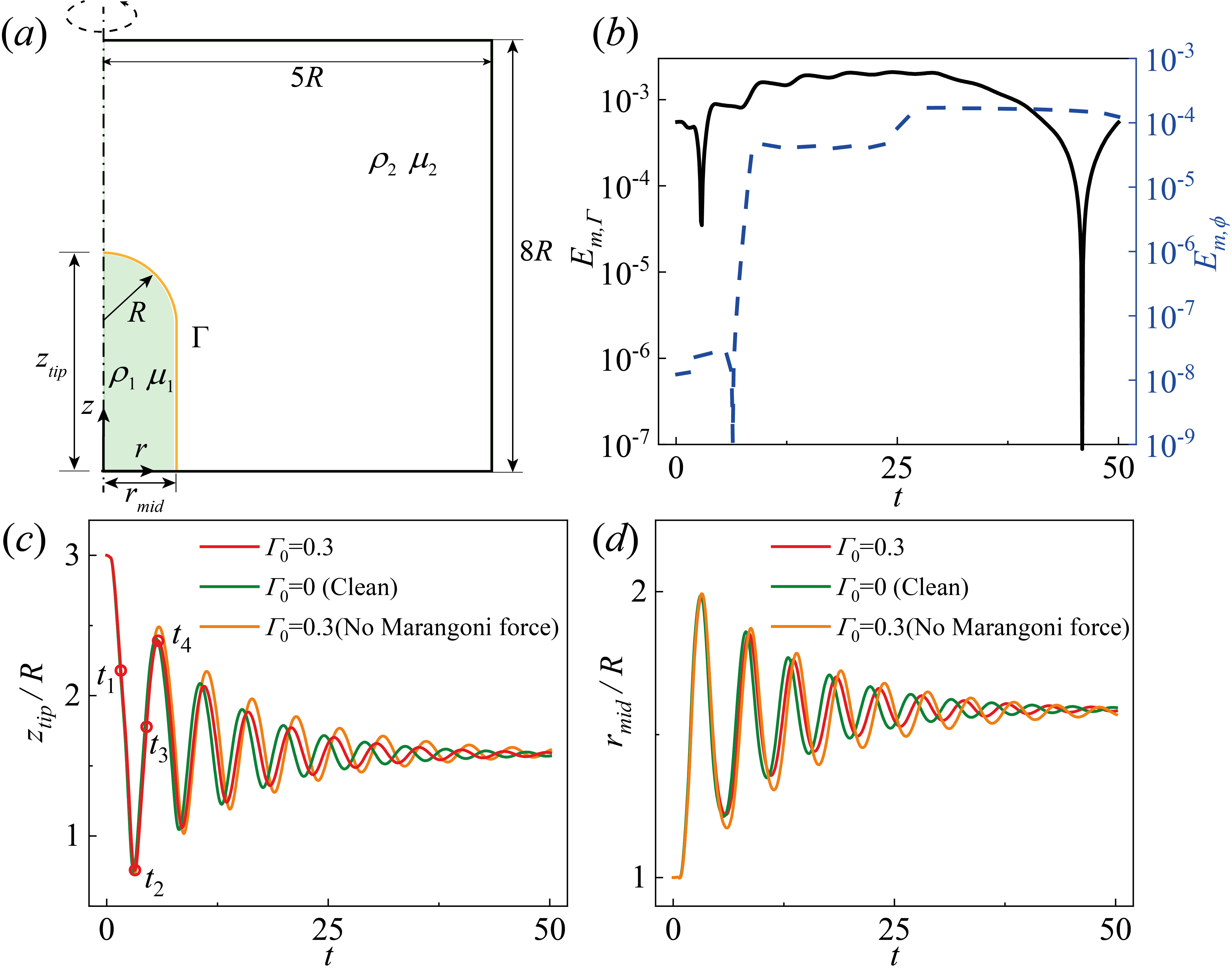}
    \caption{Contraction and oscillation of a surfactant-laden liquid filament with aspect ratio \(L/R=3\) and Ohnesorge number \(Oh=0.05\). 
    (a) Schematic of the surfactant-laden liquid filament, with \(z_{tip}\) representing the filament tip position and \(r_{mid}\) denoting the filament width at \(z=0\). 
    Initially, at \(t=0\), \(z_{tip}=3R\) and \(r_{mid}=R\). 
    (b) Temporal evolution of the mass error for surfactant concentration, \(E_{m,\varGamma}\), and for the liquid filament, \(E_{m, \phi}\). 
    (c) Time evolution of the filament tip position, \(z_{tip}\), and (d) evolution of the filament width, \(r_{mid}\). 
    Four time instants (\(t_1, t_2, t_3,\) and \(t_4\)) are defined for the first oscillation cycle of the surfactant-laden filament.
     \(t_1=t_2/2\) and \(t_3=(t_2+t_4)/2\), where \(t_2\) and \(t_4\) correspond to the moments when the filament tip reaches its minimum and maximum positions, respectively, during the first oscillation.}
    \label{fig:mesh521}
\end{figure}

We examine the contraction and oscillation behaviors of a liquid filament with finite viscosity and initially uniform surfactant concentration, as shown in Fig.~\ref{fig:mesh521}(a). 
In the absence of surfactant, a clean liquid filament with finite viscosity typically evolves into a spherical shape after experiencing a series of damped oscillations~\citep{Notz2004,wang2019}.
This behavior arises from the interaction between inertia, which drives the filament away from equilibrium, and surface tension, which works to restore the filament to its equilibrium state~\citep{wagoner2021}. 

However, the presence of surfactants introduces additional complexities to the system, as surfactant-induced Marangoni effects modify both the contraction and oscillation dynamics.
These effects arise from surface tension gradients driven by the non-uniform distribution of surfactants along the interface, impacting the filament’s evolution.
Despite the crucial role of surfactants in many practical applications, limited research has explored the dynamics of surfactant-laden filaments or drops, with the most investigations focusing on their clean counterparts~\citep{rayleigh1879capillary,wang2019}. 







We perform axisymmetric simulations of liquid filaments contraction, assuming the surrounding fluid has negligible density and viscosity. 
The computational domain is \(5R \times 8R\), as illustrated in Fig.~\ref{fig:mesh521} (a), with the origin of the cylindrical coordinates located at the filament's center.
Due to symmetry at \(z=0\) and \(r=0\), only one quarter of the filament is simulated.
Neumann boundary conditions are applied at the left (\(r=0\)) and bottom boundaries, while far-field boundary conditions are imposed on the other two sides. 
The influence of gravity is neglected in these simulations.

Considering that the dynamics of a liquid filament with finite viscosity are dominated by inertial and surface tension forces, we adopt the inertial-capillary time \(t_c=\sqrt{\rho_1 R^3/\sigma_0}\) and the filament radius \(R\) as the characteristic time and length scales.
The characteristic velocity \(U\), the Weber number \(We\), and the Reynolds number \(Re\)  are defined as \(\sqrt{\sigma_0/\rho_1 R}\), \(We=\rho_1U^2R/\sigma_0=1\), and \(Re=1/Oh=\sqrt{\rho_1 R \sigma_0}/\mu_1\), respectively. 
For the simulations, we choose an aspect ratio \(L/R=3\), an Ohnesorge number \(Oh=0.05\), and an initial surfactant concentration \(\varGamma_0=0.3\). 
The other dimensionless parameters are set as follows: P\(\acute{\text{e}}\)clet number for surfactant transport \(Pe_{\varGamma}=10\), elasticity number \(E=0.2\),  P\(\acute{\text{e}}\)clet number for the order parameter \(Pe_\phi=0.9/Cn\), interface thickness \(Cn=0.75\Delta x\),  grid resolution \(R/\Delta x=64\), time step \(\Delta t = 5\times 10^{-4}\),  density ratio \(\zeta_d=0.01\), and viscosity ratio \(\zeta_{\mu}=0.01\). 
To avoid non-physical negative values for the surface tension \(\sigma(\varGamma)\), we use the expression \(\sigma (\varGamma)=\text{max} \left(\sigma_{\epsilon}, 1+E \ln \left(1-\varGamma\right) \right) \), with a numerical threshold \(\sigma_{\epsilon}=0.05\).
This ensures stability during simulation while preserving realistic surfactant effects.

We first assess the accuracy of mass conservation for both the surfactant, denoted as \(E_{m,\varGamma}\), and filament phase, denoted as \(E_{m, \phi}\), during the contraction and oscillation, as shown in Fig.~\ref{fig:mesh521} (b). 
The results demonstrate that the total mass of surfactant are conserved throughout the deformation process, with a maximal error of approximately \(0.1\%\), while the filament phase exhibit s a maximal error of approximately $0.01\%$.
This underscores the effectiveness of the proposed numerical model in maintaining precise mass conservation for both the surfactant and filament during dynamic filament evolution.

To analyze the impact of insoluble surfactant on the contraction and oscillation dynamics of the liquid filament, we compare the time evolutions of the filament tip position \(z_{tip}\) and its middle width \(r_{mid}\) for three cases: (i) an initially uniform surfactant concentration $\Gamma_0=0.3$ (red lines), (ii) a clean filament without surfactant $\Gamma_0 = 0$ (green lines), and (iii) a filament with an initial surfactant concentration $\Gamma_0=0.3$, but with the Marangoni force disabled (orange lines), as illustrated in Figs.~\ref{fig:mesh521} (c) and (d). 
Comparing the evolution of \(z_{tip}\) and \(r_{mid}\) with and without surfactant (red vs. green lines), we observe that the presence of surfactant becomes more pronounced after the first two oscillation cycles, primarily affecting the oscillating frequency, while having a negligible influence on the amplitude.
However, an obvious difference emerges when comparing the cases with and without the Marangoni force (red vs. orange lines).
The Marangoni force reduces both the oscillation amplitude and period, indicating that surfactant-induced Marangoni stresses play a significant role in damping the filament's oscillatory dynamics.

\begin{figure}[H]
    \centering
    \includegraphics[width=1.0\textwidth]{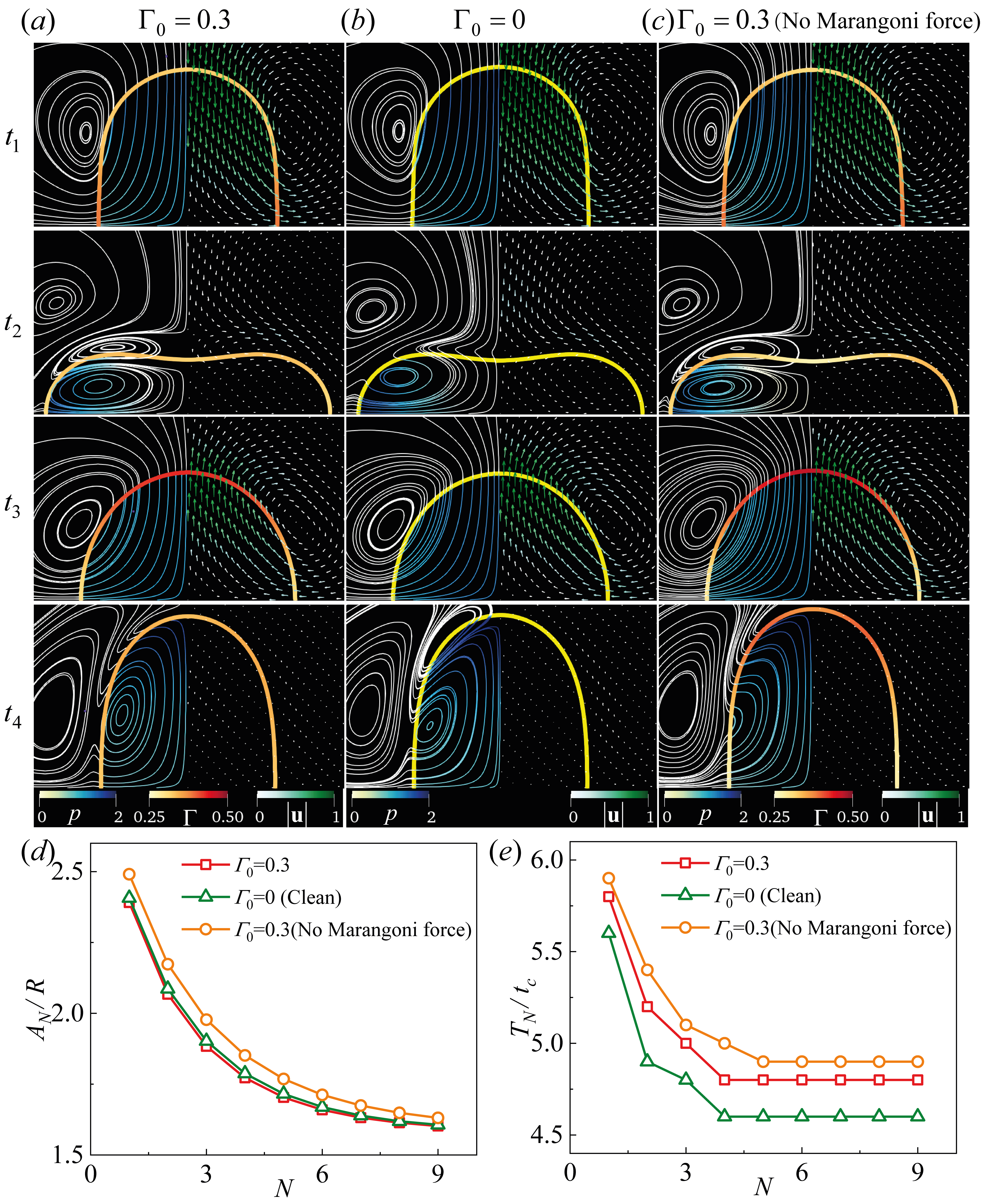}
    \caption{Numerical snapshots showing the pressure field, surfactant concentration, velocity vectors, the streamlines, and the velocity magnitude at four characteristic moments (marked by red circles in Fig.~\ref{fig:mesh521}(c)) during the first oscillation period for three cases : (a) surfactant-laden case with \(\varGamma_0=0.3\), (b) clean case with \(\varGamma_0=0\), and (c) surfactant-laden case with \(\varGamma_0=0.3\) but with the Marangoni force disabled. 
    Panels (d) and (e) display the variation in amplitude $A_N$ (normalized by filament's initial radius $R$) and period $T_N$ (normalized by inertial-capillary time $t_c$), respectively, as a function of the number of oscillation, \(N\).}
    \label{fig:mesh522}
\end{figure}

To elucidate the underlying mechanisms, we analyze the evolution of the flow field $\boldsymbol{u}$, inner pressure field $p$, and the surfactant concentration field $\varGamma$ during the first oscillation period, as illustrated in Figs.~\ref{fig:mesh522} (a)-(c). 
The filament undergoes two distinct stages within each oscillation cycle: the contraction stage (\(0< t \leq t_2\)) and the rebound stage (\(t_2 < t \leq t_4\)), where \(t_2\) and \(t_4\) correspond to the moments when the filament tip reaches its minimum and maximum positions, respectively, as shown in Figs.~\ref{fig:mesh521} (c).
 Additionally, \(t_1=t_2/2\) and \(t_3=(t_2+t_4)/2\).

During the early contraction stage (\(0\leq t \leq t_1\)), fluid is transported from the hemispherical tip toward the cylindrical region of the filament due to the capillary pressure gradient induced by the large curvature at the tip.
Simultaneously, the surfactant concentration increases as the filament interface shrinks, as seen in Fig.~\ref{fig:mesh522} (a) and (c).
By $t=t_1$, the surfactant accumulates in the middle of the filament via convection in both surfactant-laden cases, either with or without the influence of Marangoni force.
After this, the filament adopts a peanut-like shape, and the surfactant concentration becomes relatively uniform at \(t=t_2\) as depicted in Fig.~\ref{fig:mesh522} (a).

During the contraction stage (\(0\leq t \leq t_2\)), the small gradient in surfactant concentration results in a minimal Marangoni effect on filament contraction. 
The effective Ohnesorge number can be calculated as \(Oh_e=\sqrt{\sigma_0/\sigma_e} Oh=1.148Oh\), where \(\sigma_e\) is the effective surface tension coefficient based on the average surfactant concentration. 
The value of \(Oh_e\) for the surfactant-laden case is nearly identical to that of the clean case, indicating that the presence of surfactant has a little impact  on the dynamics of filament contraction.

After the contraction stage, the filament begins to rebound due to the pressure gradient in radial direction, which is accompanied by the accumulation of surfactant at filament tip as a result of convection at \(t=t_3\), as shown in Fig.~\ref{fig:mesh522} (a) and (c). 
At this point, the surfactant concentration exhibits a steeper gradient compared to the contraction stage, leading to an increase in the magnitude of the Marangoni force. 
Subsequently, this Marangoni force weakens the upward flow along the interface and inhibits the filament's rebound. 
By \(t=t_4\), the flow even reverses, reducing the surfactant concentration gradient, as depicted in Fig.~\ref{fig:mesh522} (a). 
In cases where the Marangoni force is neglected, the tangential stretching of the filament interface similarly reduces the surfactant concentration gradient at \(t=t_4\) as shown in Fig.~\ref{fig:mesh522} (c).
During the rebound stage, the surfactant concentration gradient induces a Marangoni force, which in turn diminishes the concentration gradient by inducing a reverse flow. 






 Fig.~\ref{fig:mesh522} (d) and (e) show the variation in oscillation amplitude \(A_N\) and period \(T_N\) as a function of the number of oscillation, \(N\), where \(A_N\) is defined as \(A_N = z_{tip}\) at the moment when the filament completes an oscillation.
When the filament reaches a steady state, it evolves into a spherical shape with a radius of \(R_s=1.587\), which corresponds to the initial volume of filament \(S_0=8/3\pi R^3\). 
Both the filament tip position, \(z_{tip}\), and the midpoint radius, \(r_{mid}\), eventually converge to this value.
For all three cases, the oscillation amplitude steadily decreases as  \(N\) increases, primarily due to viscous damping.
Notably, the oscillation period decreases up until the fourth oscillation cycle (\(N\leq4\)) and then remains constant thereafter, which holds true across all  cases.
This behavior aligns with the theoretical predictions for the oscillation of a clean drop, as described by Lamb~\citep{lamb1881oscillations,lamb1924hydrodynamics}, which asserts that the oscillation period remains constant for finite-amplitude oscillations.
 This predication also holds for surfactant-laden drops.
 One possible interpretation is that the capillary force dominates over the Marangoni force once the surfactant concentration becomes approximately uniform for \(N>4\). 
 As a result, with increasing oscillation cycles \(N\), the presence of surfactant primarily influences the filament's oscillation by modulating the capillary force.





\section{Conclusion}\label{sec:s6}


 In conclusion, we have developed a numerical model to simulate two-phase flow with insoluble surfactant transport along the interface, where surfactant mass conservation is enhanced. 
 To carefully capture the two-phase interface, we employ the profile-preserving approach~\citep{HAO2024104750}, which ensures the equilibrium state of the diffuse interface profile. 
 The surfactant concentration is modeled using a diffuse-interface formulation of the surfactant transport equation within the framework of the advective Chan-Hilliard model, relying on a regularized delta function for accurate representation.
To solve the surfactant transport equation, we use a semi-implicit scheme, allowing us to avoid stringent time step restrictions.
Spatial discretization is handled using the finite volume method, ensuring a conservative formation. 
Furthermore, we introduce a hybrid method (FECSF model) for modeling the surface tension force, where the capillary and Marangoni forces are computed using the interfacial free-energy density model and the continuum-surface force (CSF) model, respectively. This approach minimizes numerical errors in the computation of surface tension, especially in two-phase flow scenarios with large density and viscosity ratios.
This refined model improves accuracy and stability for complex two-phase flow simulations involving surfactant transport.
 
 



 We then conduct three verification tests including the diffusion of surfactant on both a stationary and an expanding circle, as well as the assessment of spurious currents on a drop. 
 The results demonstrate the performance of the present model in terms of surfactant mass conservation, mesh convergence, and numerical accuracy. 
 We preform a simulation of a 2D drop with insoluble surfactant in a shear flow, which shows quantitative agreement with previous studies~\citep{XU20125897}. 
 Additionally, we investigate the contraction and oscillation dynamics of a surfactant-laden liquid filament. 
 Our findings indicate that the Marangoni force, arising from the surfactant concentration gradient, can suppress the oscillation amplitude—particularly during the rebound stage—thereby reducing the gradient by inducing reverse flow. 
 The presence of surfactant also leads to a reduction in capillary force, which in turn increases the oscillation period.

 In contrast to the dynamics of insoluble surfactants, soluble surfactants are advected and diffused both at the interface and within the bulk fluid, accompanied by mass transfer processes (i.e., adsorption and desorption) between the interface and bulk fluid~\citep{Manikantan2020}. 
However, it is important to note that when the timescale associated with surfactant adsorption and desorption significantly exceeds that of the underlying flow dynamics, soluble surfactants exhibit behavior similar to that of purely insoluble surfactants~~\citep{Kamat2020}. 
In such cases, soluble surfactants can be approximated as insoluble surfactants by neglecting their advection and diffusion in the bulk fluid as well as their mass transfer effects. 
Consequently, the present model can effectively simulate this type of flow.


In future work, we plan to extend the numerical model to two-phase flow with soluble surfactants by incorporating the surfactant transport equation in the bulk fluid, along with the adsorption and desorption processes. 
Additionally, we will develop a three-dimensional model utilizing an adaptive mesh refinement (AMR) method to achieve localized refinement of the interface. 
Those enhancements aim to improve the accuracy and efficiency of simulations involving complex surfactant dynamics.

\section*{Acknowledgements}
This work has received financial support from the Natural Science Foundation of China (Grant No. 12102171), the Natural Science Foundation of Shenzhen (Grant No. 20220814180959001), and the Guangdong Basic and Applied Basic Research Foundation (Grant No. 2024A1515010509 and 2024A1515010614).

\appendix

\section{Sharp-interface and diffuse-interface formulations of surfactant transport equation}
\label{sec:sample:appendix}

The surfactant concentration is gorvened by the time-dependent advection-diffusion equation~\citep{Scriven1960,stone1990,wong1996} and its dimensionless form in sharp-interface formulation can be written as 
\begin{equation} \label{STE_O1}
    \ \frac{\partial \varGamma }{\partial t}+{{\nabla }_{s}}\cdot \left( \varGamma {{\mathbf{u}}_{s}} \right)+\varGamma \left( {{\nabla }_{s}}\cdot \mathbf{n} \right)(\mathbf{u}\cdot \mathbf{n})=\frac{1}{Pe_{\varGamma}} \nabla _{s}^{2}\varGamma,\ 
\end{equation}
where  \(\varGamma=\varGamma^\ast /\varGamma_{\infty}^\ast\) represents the dimensionless surfactant concentration at the interface, \(\varGamma_{\infty}^\ast\) is its saturation interfacial concentration, \(\nabla_s={(\mathbf{I-nn})\cdot \nabla}\) is the surface gradient operator, \(\mathbf{u}\) is the dimensionless fluid velocity, \(\mathbf{u_s}=(\mathbf{I-nn})\cdot \mathbf{u}\) denotes the tangential velocity along the interface, and \(\mathbf{n}\) is the normal direction of the interface. \(Pe_{\varGamma}=UR/D\) is the P\(\acute{\text{e}}\)clet number representing the rate ratio between advective and diffusive surfactant transport, where \(D\) is the diffusion coefficient of surfactant, \(U\) is the characteristic velocity, and \(R\) is the characteristic length. 


The surfactant transport equation in sharp-interface formulation can be rewritten in a diffuse-interface one, as~\citep{ERIKTEIGEN2011375}
\begin{equation} \label{STE_O2}
    \frac{\partial \left( {{\delta }}\varGamma  \right)}{\partial t}+\nabla \cdot \left( {{\delta }}\varGamma \mathbf{u} \right)=\frac{1}{Pe_{\varGamma}} \nabla \cdot \left( {{\delta }}\nabla \varGamma  \right),  
\end{equation}
where the surfactant concentration \(\varGamma \) defined on the interface \(S\) has been extended to the general domain \(\varOmega \) by introducing a surface delta function \(\delta\), written as~\citep{ERIKTEIGEN2011375}
\begin{equation} \label{STE_a2}
    \int_{S} \varGamma \mathrm{~d} S=\int_{\varOmega} \varGamma \delta \mathrm{~d} \varOmega = \int_{\varOmega} \hat{\varGamma } \mathrm{~d} \varOmega.
\end{equation}

We refer the readers to~\citep{ERIKTEIGEN2011375} for the derivation in detail.

 \bibliographystyle{elsarticle-num} 
 \bibliography{cas-refs}





\end{document}